\documentclass[preprint,superscriptaddress,nofootinbib]{revtex4-1}
\usepackage{hyperref}
\usepackage{bbm}
\usepackage{amsfonts}
\usepackage{mathrsfs}

\usepackage{amsmath,amssymb}

\usepackage{epsfig}
\usepackage{graphicx}               % Standard graphics package
\usepackage{url}
\usepackage{hyperref}
\usepackage{float}
\usepackage{pstricks}
\usepackage{color}
\usepackage{multirow}

\newcommand{\nc}{\newcommand}

\nc{\beq}{\begin{equation}}
\nc{\eeq}{\end{equation}}
\nc{\barray}{\begin{eqnarray}}
\nc{\earray}{\end{eqnarray}}
\nc{\barrayn}{\begin{eqnarray*}}
\nc{\earrayn}{\end{eqnarray*}}
\nc{\bcenter}{\begin{center}}
\nc{\ecenter}{\end{center}}
\nc{\mc}{\mathcal}
\nc{\er}[1]{(\ref{eq:#1})} 
\nc{\onehalf}{\frac{1}{2}}
\nc{\partialbar}{\bar{\partial}}
\nc{\psit}{\widetilde{\psi}} 
\nc{\Tr}{\mbox{Tr}}
\nc{\hc}{\mbox{H.c.}}
\nc{\ev}{\;\mathrm{eV}}
\nc{\mev}{\;\mathrm{MeV}}
\nc{\gev}{\;\mathrm{GeV}}
\nc{\tev}{\;\mathrm{TeV}}
\nc{\f}{\frac}

\def\chii0{\chi_i^0}
\def\chij0{\chi_j^0}

\newcommand{\gsim}{\lower.7ex\hbox{$\;\stackrel{\textstyle>}{\sim}\;$}}
\newcommand{\lsim}{\lower.7ex\hbox{$\;\stackrel{\textstyle<}{\sim}\;$}}
\nc{\ttbar}{t\bar t}

\begin{document}

\begin{flushright}
WSU-HEP-1508
\end{flushright}

\title{Closing the Wedge: Search Strategies for Extended Higgs Sectors with Heavy Flavor Final States}

\author{Stefania Gori}
\affiliation{Perimeter Institute for Theoretical Physics, 31 Caroline St. N, Waterloo, Ontario, Canada}
\affiliation{Department of Physics, University of Cincinnati, Cincinnati, Ohio 45221, USA}
\author{Ian-Woo Kim}
\affiliation{CERN TH-PH Division, Meyrin, Switzerland}
\affiliation{UpHere, San Francisco, CA 94122}
\author{Nausheen R. Shah}
\affiliation{Department of Physics and Astronomy, Wayne State University, Detroit, MI 48201, USA}\author{Kathryn M. Zurek}
\affiliation{Theoretical Physics Group, Lawrence Berkeley National Laboratory, Berkeley, CA 94720, USA  \\  Berkeley Center for Theoretical Physics, University of California, Berkeley, CA 94720}

\vskip 10 pt

\begin{abstract}

We consider search strategies for an extended Higgs sector at the high-luminosity LHC14 utilizing multi-top final states.  In the framework of a Two Higgs Doublet Model, the purely top final states ($t\bar t, \, 4t$) are important channels for heavy Higgs bosons with masses in the wedge  above $2\,m_t$ and at low values of $\tan\beta$, while a $2 b 2t$ 
final state is most relevant at moderate values of  $\tan \beta$.  We find, in the $t\bar t H$ channel, with $H \rightarrow t \bar t$, that both single and 3 lepton final states can provide statistically significant constraints at low values of $\tan \beta$ for $m_A$ as high as $\sim 750$ GeV.  When systematics on the $t \bar t$ background are taken into account, however, the 3 lepton final state is more powerful, though the precise constraint depends fairly sensitively on lepton fake rates.  We also find that neither $2b2t$ nor $t \bar t$ final states provide constraints on additional heavy Higgs bosons with couplings to tops smaller than the top Yukawa due to expected systematic uncertainties in the $t \bar t$ background.

\end{abstract}

\maketitle

\tableofcontents

%%%%%%%%%%%%%%%%%
\section{Introduction}\label{sec:intro}
%%%%%%%%%%%%%%%%%

Precision measurements of the observed 125 GeV Higgs boson, as well as  searches for new Higgs bosons, are a focus of Run II of the LHC.  These measurements will give us crucial insights into the nature of Electroweak Symmetry Breaking (EWSB).  This program, for the case of the 125 GeV Higgs, is already well-defined and under investigation. Current measurements have large uncertainties, but overall consistency with SM expectations is observed. Both ATLAS and CMS expect to reduce their uncertainties over the next few years, achieving percent level precision for decays into massive gauge bosons and  $\sim 5\%$ level precision for decays into heavy flavor fermions \cite{ATL-PHYS-PUB-2014-016,CMS:2013xfa}. 
        
If additional Higgs bosons are present, the so-far SM-like nature of the observed 125 GeV Higgs boson implies that such states can only be minimally mixed. This suggests that the Higgs sector is approximately aligned~\cite{Gunion:2002zf,Delgado:2013zfa,Craig:2013hca,Asner:2013psa,Carena:2013ooa,Haber:2013mia, Dev:2014yca,Carena:2014nza,Das:2015mwa,Dev:2015bta,Bernon:2015qea,Dev:2015cba,Haber:2015pua,Carena:2015moc}, either via decoupling or via alignment without decoupling. In the latter scenario, it becomes essential to directly search for the presence of extra Higgs bosons, which may have an unmeasurably small impact on the properties of the 125 GeV Higgs. Thus the program of precision Higgs measurements is complimentary to direct searches for electroweak (EW) scale Higgs bosons. 

To date, direct searches for heavy Higgs bosons (both scalar $H$ and pseudoscalar $A$) have focused on decays to bottom quarks and $\tau$-leptons, which dominate the branching fraction at large $\tan\beta$ ($t_\beta$) in a Type II Two Higgs Doublet Model~(2HDM).  The most commonly studied channels are Higgs production in association with bottom quarks, which is enhanced by $\tan\beta$, as well as gluon fusion.  The $(b\bar b) H/A \rightarrow b\bar{b}\tau^+\tau^-$ channel gives the strongest constraint, with the bound $t_\beta\lesssim 45$ for a heavy Higgs with mass below 1 TeV~\cite{ATLAStautau13}, obtained using 3.2 fb$^{-1}$ of 13 TeV data (see the red region in Fig.~\ref{fig:summary}).  The constraint arising from the search channel $b\bar b H/A \rightarrow b\bar b b\bar b$~\cite{Chatrchyan:2013qga,Malone:2015mia} is weaker in the context of a Type II 2HDM, but nevertheless very important  in more general 2HDMs, for which the coupling of $H/A$ to $\tau$-leptons and bottom-quarks is independent~\cite{Carena:2012rw}.\footnote{One example is the aligned 2HDM studied for example in Ref.~\cite{Altmannshofer:2012ar}.}

 There are no direct constraints at low and moderate values of $\tan\beta$, and for heavy Higgs masses above the top threshold for Type II models.  Additionally, projections of the present ATLAS and CMS analyses show that part of this region of parameter space will remain unexplored at the High Luminosity Run~\cite{Lewis:2013fua,Baglio:2015wcg}. Here, for $t_\beta\lesssim 5$ and $m_{H/A}\geq 350$ GeV, the main decay mode is $H/A\to t\bar t$, for which there are no experimental analyses to date.

The purpose of this paper is to study search strategies for additional Higgs bosons at low and moderate values of $\tan \beta$.  We study $b\bar b H/A \rightarrow b\bar b t\bar t$~($2b2t$), $t\bar t H/A \rightarrow  t\bar t b\bar b$~($2t2b$) (both important at moderate $t_\beta$), $t\bar t H/A \rightarrow t \bar t t\bar t$~($4t$) and $H/A \rightarrow t \bar t$~($2t$) (both important at low $t_\beta$).

Previous studies have also focused on constraining heavy Higgs bosons through decays to top quarks at the LHC \cite{Craig:2015jba,Hajer:2015gka}.   Our study differs in several aspects, drawing new conclusions in several instances.  Ref.~\cite{Craig:2015jba} considered the $4 t$ final state, though this study implemented and scaled the existing CMS multi-lepton analyses.  We instead design two possible search strategies (one lepton + $b$ jets and three leptons + $b$ jets), estimating the impact of future lepton fake rates, as well as $b$ tagging efficiencies, on our conclusions. We also discuss in detail the relevance of systematic effects in the one lepton + $b$ jets channel. As a result of this analysis, we find a more significant reach than Ref.~\cite{Craig:2015jba}.  This reference also showed that systematics fundamentally limit the reach of the $ H \rightarrow t \bar t$ channel.  We confirm this conclusion, though our methods significantly differ in that, for the first time, we have implemented in a Montecarlo tool, {\tt MadGraph}, the interference between the heavy Higgs signal and the SM background, reconstructing the fully interfered signal and background after showering and detector effects.  Ref.~\cite{Hajer:2015gka}  considered the $2b2t$ final state, though this study, based on a Boosted Decision Tree method, found a greater impact of forward $b$-jets on the significance of the signal, as the $b$-quark distributions of this study differed from ours. In our cut-based analysis, we also find that systematic uncertainties severely limit the reach of the $2b2t$ final state.  As a result, whereas Ref.~\cite{Hajer:2015gka} found significant bounds on heavy Higgs bosons in the $2b2t$ final state, we find that it will be difficult to draw constraints.

The outline of this paper is as follows.  In the next section we review rates and signatures for the channels we study.  We summarize the main features that we will utilize in our analyses.  Then we move systematically through the three final states we consider in detail: $4t$ (Sec.~\ref{sec:4t}), $2b2t$ (Sec.~\ref{sec:2t2b}) and $t \bar t$ (Sec.~\ref{sec:2t}).  We highlight the challenges and advantages of each search, and suggest ways that experiments could improve the reach in each case.  Finally, we conclude. In Appendix A we discuss our top reconstruction method.

\section{Rates and Signatures from Aligned Heavy Higgs Bosons}\label{sec:ProdDecay}

\begin{figure}[tbph]
\begin{center}
\includegraphics[width=0.45 \textwidth]{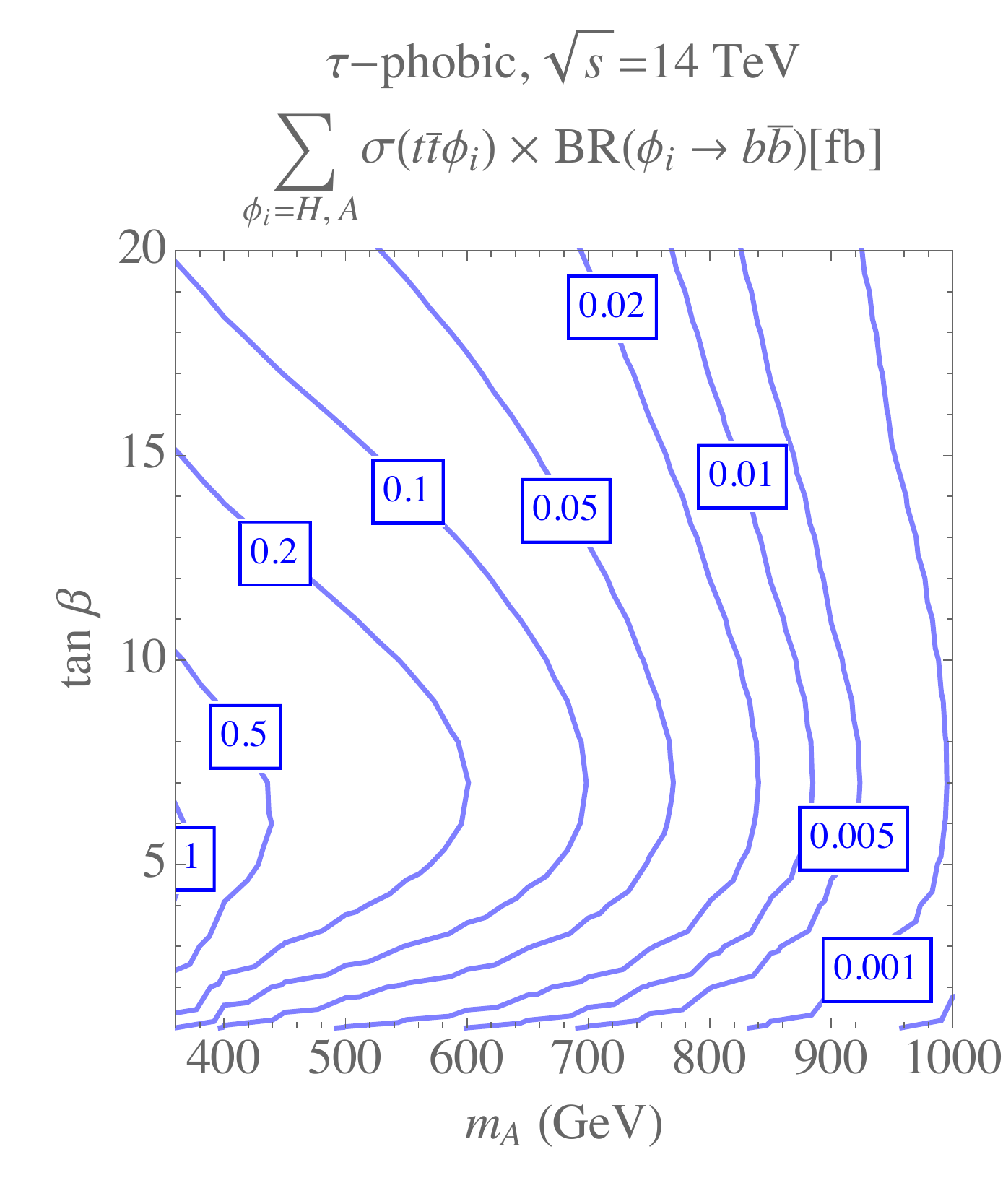} \includegraphics[width=0.45\textwidth]{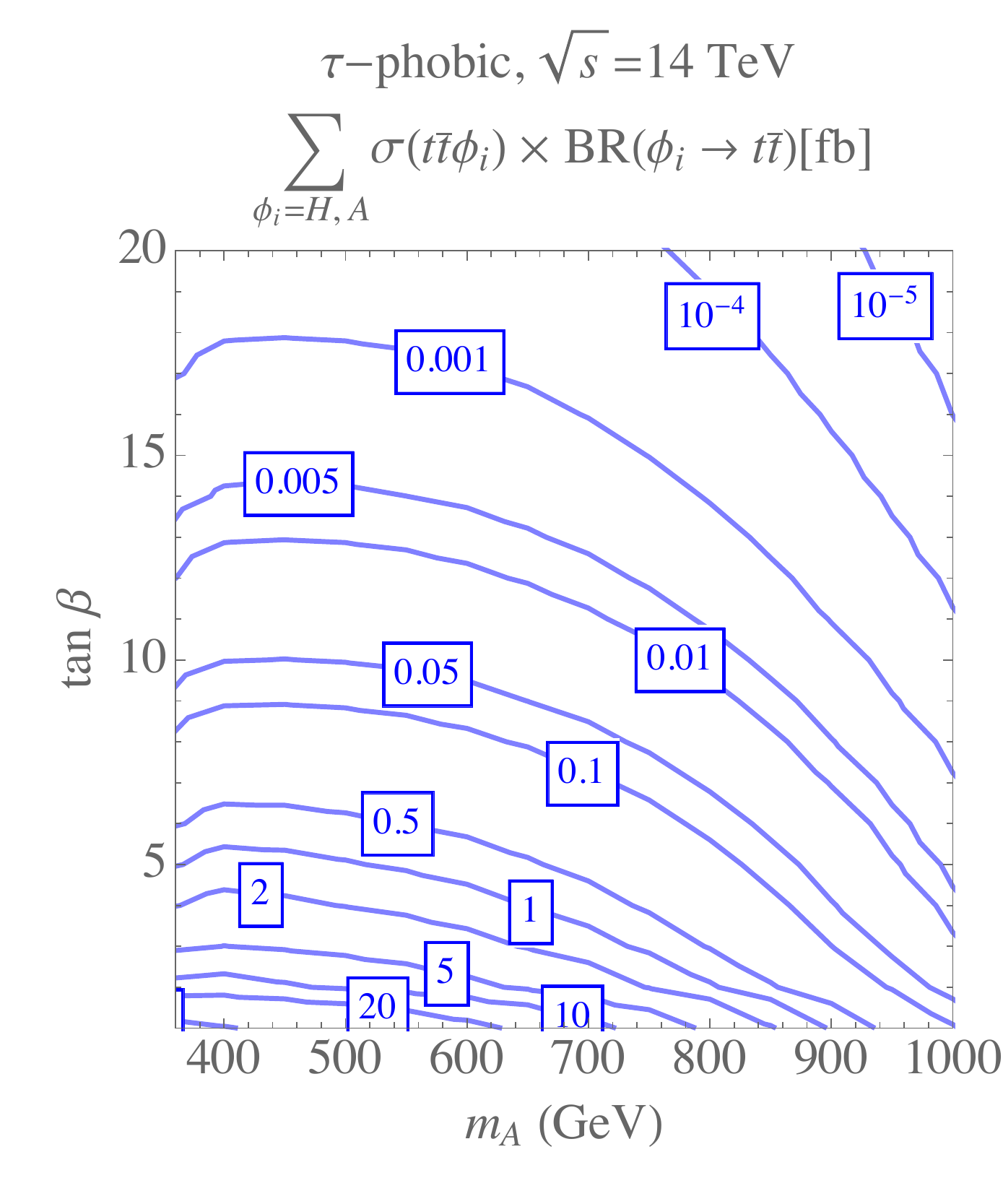}\\
\includegraphics[width=0.45\textwidth]{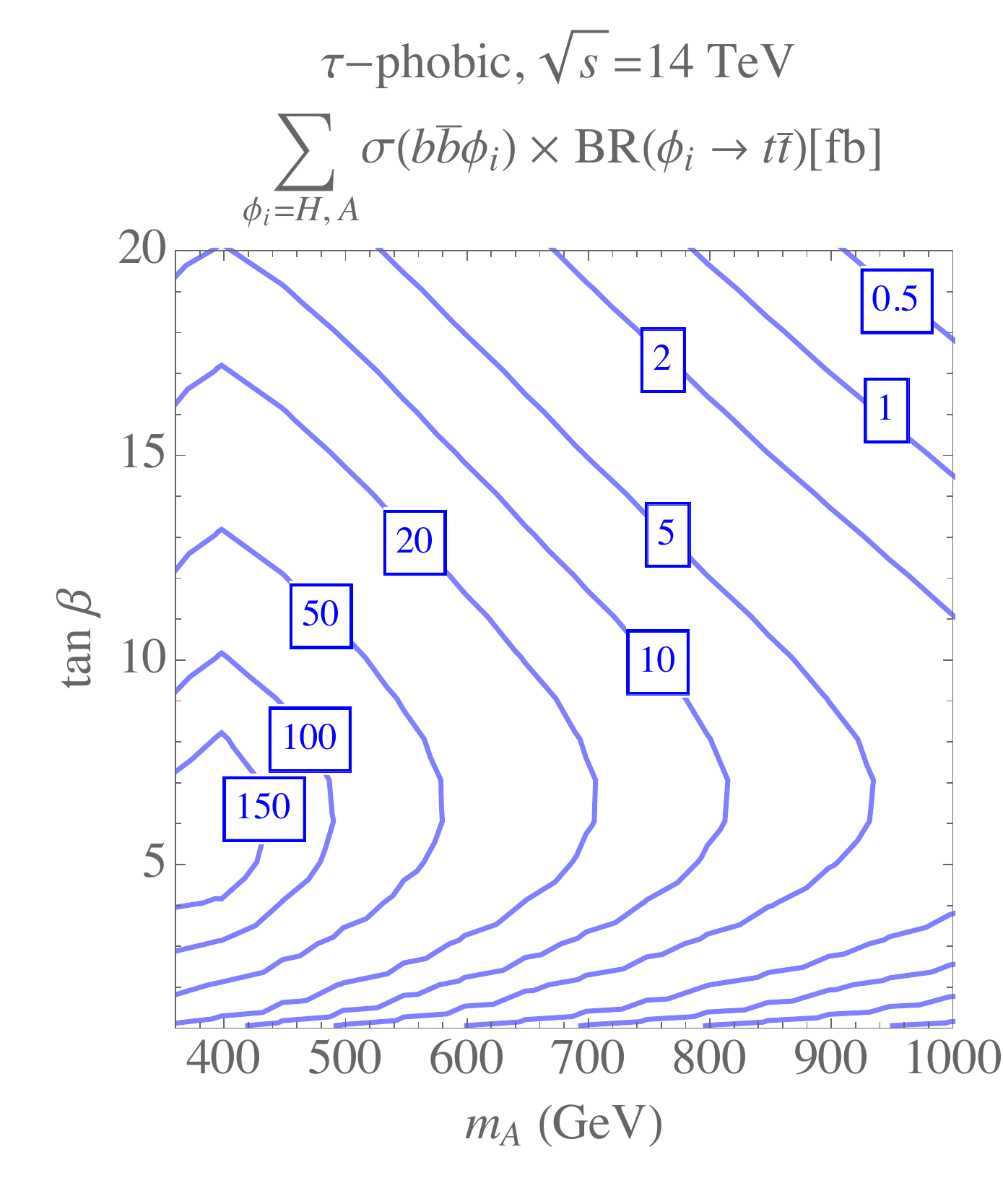}\includegraphics[width=0.45 \textwidth]{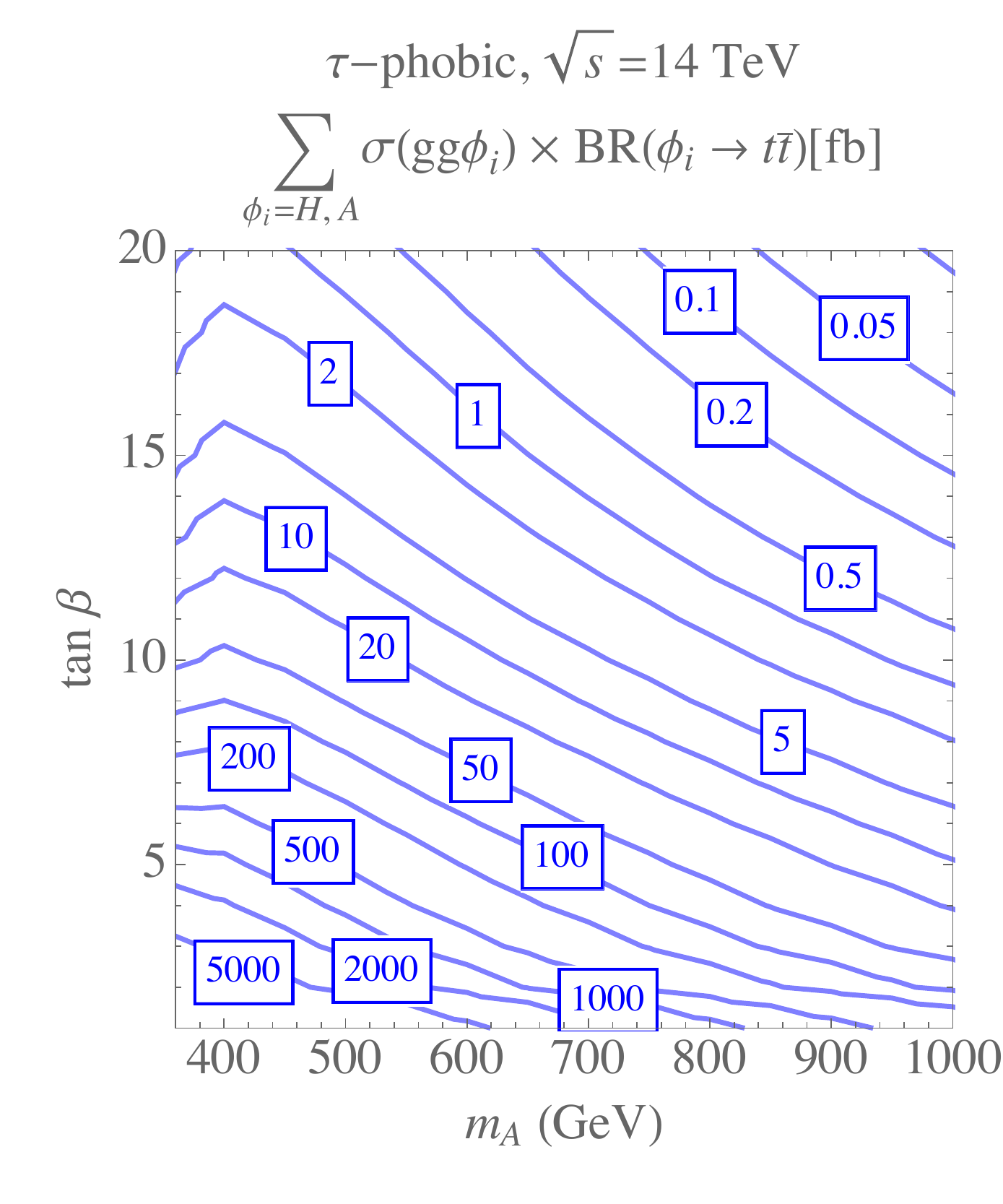} 
\end{center}
\caption{Production cross-section times branching fraction for the channels of interest. For concreteness, we adopt the $\tau$-phobic MSSM scenario~\cite{Carena:2013ytb}. }
\label{fig:BRttandSigma}
\end{figure}

We begin by summarizing production rates and branching fractions for heavy scalars in a Type II 2HDM. A Type II 2HDM consists of doublets $H_u$, coupling only to up-type quarks, and $H_d$, coupling only to down-type quarks and leptons. The neutral scalar components of both these fields acquire vacuum expectation values, $v_u$ and $v_d$, with $v_u^2+v_d^2=v^2=(246~\rm{GeV})^2$ and $\tan\beta=v_u/v_d$. Such a 2HDM gives rise to the physical states: $h$ and $H$ (scalars), $A$ (pseudoscalar) and $H^\pm$ (charged Higgses).

For simplicity, we assume that we are in the (almost) alignment limit~\cite{Gunion:2002zf,Delgado:2013zfa,Craig:2013hca,Asner:2013psa,Carena:2013ooa,Haber:2013mia, Dev:2014yca,Carena:2014nza,Das:2015mwa,Dev:2015bta,Bernon:2015qea,Dev:2015cba,Haber:2015pua,Carena:2015moc}, in which, to first approximation, the mass eigenstates are given by the fields in the {\it Higgs} basis~\cite{Georgi:1978ri,Donoghue:1978cj,Lavoura:1994fv,Lavoura:1994yu,Botella:1994cs,Branco:1999fs}, where only one doublet field combination~($H_{SM}$) couples to gauge bosons and does not mix with the second orthogonal doublet combination~($H_{NSM}$). 
The observed Higgs boson~($h$) with a mass of a 125 GeV is identified with the scalar arising from $H_{SM}$, whereas the $H_{NSM}$ gives rise to $H$, $A$ and the charged Higgs bosons. In particular, alignment implies that the couplings of $H$ to $W^+W^-/ZZ/hh$ go to 0, as does the coupling of $A$ to $Zh$. It should be noted that in the almost alignment limit, below the top decay threshold ($m_{H/A} \lesssim$ 350 GeV) at low values of $\tan\beta$, both $H$ and $A$ could have significant decays into these channels in spite of the suppressed couplings, due to the paucity of other available decay channels. The coupling between $AZH$ is not suppressed, though in a typical 2HDM spectrum where $A$ and $H$ are almost degenerate,
this decay is kinematically forbidden.\footnote{In a generic 2HDM arising for example from non-minimal SUSY models like the Next to Minimal Supersymmetric SM~(NMSSM), where such a degeneracy in the $H$ and $A$ mass is not necessarily expected, the decay of $A\to ZH$ or $H\to Z A$ can be very relevant at low values of $\tan\beta$, even after the top threshold opens up~(see for eg. Refs.~\cite{Coleppa:2014hxa, Dorsch:2014qja, Carena:2015moc, CMS-PAS-HIG-15-001,Dorsch:2016tab}).}  In contrast to the above decay channels induced by tree-level couplings, both $A$ and $H$ couple to pairs of gluons and photons only via loops of colored/charged particle as does the SM-like Higgs. Generically, the branching ratio into pairs of gluons or photons is expected to be $\mathcal{O}(10^{-3})$ or $\mathcal{O}(10^{-5})$ respectively. 

In the alignment limit, the expressions for the couplings of the new Higgs boson with fermions are rather simple: $H t\bar t\sim m_t/(v \,t_\beta)$, $H b\bar b\sim m_b t_\beta/v $ and $H \tau\bar \tau\sim m_\tau t_\beta/v $. In the alignment limit, these couplings correspond also to the couplings of the pseudoscalar $A$ to fermions.  Corrections away from the alignment limit scale like $x\; m_q/v$, where $x$ quantifies the deviations from perfect alignment, with $\alpha=\beta-\pi/2+x$, and $\alpha$ the mixing angle between the  Higgs doublets $H_u$ and $H_d$. From the precision measurements of the 125 GeV Higgs boson, $|x|$ is expected to be at most $\sim 0.1$ depending on $t_\beta$.
Hence, at low values of $t_\beta$, the couplings of the non-standard Higgs bosons to the top quark are sizable, leading to large production cross-sections from gluon fusion and the expectation of large branching fraction into top quark pairs when kinematically allowed. The production of $H$ or $A$ in association with a pair of top quarks is however not expected to be particularly large at a centre of mass energy of 14 TeV, due to the large top quark mass (see top row of Fig.~\ref{fig:BRttandSigma}). Since the coupling to $b$ quarks is enhanced by $t_\beta$, at moderate values of $t_\beta$ the gluon fusion production receives comparable contributions from the top and bottom loops,  with approximately equal contributions from both loops at $t_\beta\sim \sqrt{m_t/m_b}\sim6$.  The associated production with $b$ quarks is expected to increase with $t_\beta$. 

For concreteness, in Fig.~\ref{fig:BRttandSigma}, we compute the production cross-sections times branching ratios of the non-standard Higgs bosons using the latest version of {\tt FeynHiggs2.11.2}~\cite{Hahn:2013ria} for the $\tau$-phobic Minimal Supersymmetric SM~(MSSM) scenario~\cite{Carena:2013ytb}, with heavier sleptons $M_{\tilde{l}_{1,2,3}}=1$ TeV. We list the relevant MSSM parameters here for convenience:
\begin{eqnarray}\nonumber
&&M_{SUSY}=1500 \; {\rm{ GeV}},
\mu = 2000 \; {\rm{ GeV}},
X_t^{\rm OS} = 2.45 \;M_{ {\rm{SUSY}}},
A_b=A_\tau=A_t,
M_1=100 \; {\rm{ GeV}},\\
&&M_2=200 \; {\rm{ GeV}},
M_3=1500 \; {\rm{ GeV}},
M_{\tilde{q}_{1,2}} = 1500\; {\rm{ GeV}},
A_f=0 \quad (f = c,s,u,d,\mu,e).
\end{eqnarray}
For this figure, we sum the contributions of the scalar $H$ and the pseudoscalar $A$, since the mass splitting is always below $\sim (10-20)$ GeV, and hence below the
expected
experimental mass resolution.
This scenario at $\tan\beta\sim 10$ is very well approximated by a Type II 2HDM in the alignment limit. Note that for values of $\tan\beta$ much larger or much smaller than 10, the $H t\bar{t}$ coupling can deviate from the alignment value by $\mathcal{O}(10\%)$, depending on the precise value of $m_A$. However, this effect will mainly be canceled in signal rates for $H$ in the region of parameter space of interest. The effect on the coupling of $Hb\bar{b}$ is in contrast suppressed by $t_\beta$ and therefore  completely irrelevant. 
Hence gluon fusion at low values of $t_\beta$ and the associated top production channels in the alignment limit are very similar to the numbers shown in Fig.~\ref{fig:BRttandSigma}.

In the figure, $\tan\beta$ is taken in the [1-20] interval, as well as $m_A\in[400,1000]\,$ GeV. This region of parameter space is rather unexplored by experimental searches to date.
Other regions are, instead, already probed or will be probed in the coming years by the LHC.   At larger values of $\tan\beta$, as mentioned in the Introduction, constraints are derived from heavy Higgs decaying into $\tau$-leptons and $b$-quarks.  These searches are not yet able to probe the $m_A\geq 400$ GeV and $\tan\beta\leq 20$ region, presented in Fig.~\ref{fig:BRttandSigma}.  This remains true at the High Luminosity LHC~\cite{Lewis:2013fua,Baglio:2015wcg}.   In principle, decays of heavy Higgs bosons to gauge bosons and the SM Higgs may be highly constraining, especially below the top threshold ($m_{H,A} \lesssim 350$ GeV) where the branching fraction may easily be large (see, for example Ref.~\cite{Djouadi:2015jea}). These constraints depend strongly, however, on the degree to which the Higgs sector is aligned.

In this paper we focus on the ``wedge'' of open parameter space with $t_\beta \lesssim 10$ and $m_{H,A}\gtrsim 350$ GeV, where the Higgs bosons decay dominantly to tops, and the $b \bar b$/ $\tau^+\tau^-$ final states are not constraining.  As a result, we consider $t \bar t$, $4t$ and $2t2b/2b2t$ signatures.  From Fig.~\ref{fig:BRttandSigma} we see that the rates for associated top production processes ($4t$ and $2t2b$) are very suppressed compared to either gluon fusion initiated~($2t$) or associated $b$-quark production ($2b2t$), both of which reach 100's of fb. In particular, the $4t$ signature may be as large as 10's of fb at low values of $\tan\beta$. In spite of the smaller rates, since this signature is quite spectacular, we expect that it can offer some reach in the $\tan\beta-m_A$ plane (see Sec.~\ref{sec:4t}). On the other hand, the $2t2b$ signature ($pp\to t\bar t H/A, H/A\to b\bar b$) is always at least two orders of magnitude smaller in rate than the corresponding $2b2t$ ($pp\to b\bar b H/A, H/A\to t\bar t$; note that we differentiate between $2t2b$ and $2b2t$). Since we do not expect that the reconstruction of the $H/A$ resonance from $b$ quarks will bring much improved significance compared to the reconstruction from top quarks, we will not discuss this signature further, instead focusing on the $2b2t$ signature in Sec.~\ref{sec:2t2b}.\footnote{Note that, with the analysis performed in Sec.~\ref{sec:2t2b} for the $2b2t$ signature, the $2t2b$ signature is also included as part of the simulation. However, we have checked that, in spite of the relatively good acceptance, this signal amounts at most to $\mathcal O(5\%)$ of the $2b2t$ signal in the entire region of parameter space of interest, for the optimal cuts we find in Sec.~\ref{sec:2t2b}.}

Note that this is a unique time to test heavy Higgs bosons produced in association with top quarks. In fact, at the 8 TeV LHC, the corresponding $t\bar t H/A$ production cross section is at most $\mathcal O(1\, {\rm{ fb}})$ for $m_H=400$ GeV and low values of $\tan\beta$, at least an order of magnitude smaller than the cross section at the 14 TeV LHC (see upper right panel of Fig.~\ref{fig:BRttandSigma}).

%%%%%%%%%%%%%%%%%
\section{Heavy Higgs Signals in Heavy Flavor Final States at LHC14}\label{sec:collider}
%%%%%%%%%%%%%%%%%

We model signals and backgrounds using {\tt MadGraph5}~\cite{Alwall:2014hca}, interfaced with {\tt Pythia 6.4}~\cite{Sjostrand:2006za} for parton showering. For detector simulation, we modify {\tt PGS4}~\cite{PGS} to enable anti-kT jet reconstruction. We generate events for the $t\bar t$ background, allowing up to two additional partons in the final state, and adopt the MLM matching scheme~\cite{Mangano:2006rw}, with $xqcut$ = 20 GeV.  We also generated $W(Z)~+$ jets, $t\bar t W$, $t\bar t Z$ and 4 top backgrounds. As we will discuss in Sec.~\ref{sec:2t2b}, we generate signal and background samples with the four flavor scheme, and we have checked that, with the specific analyses performed, signal rates are not modified by more than a few percent if we use the five flavor scheme. We use {\tt SUSYHIT} \cite{Djouadi:2006bz} to generate the several SUSY benchmark scenarios $m_A=(400-1000)$ GeV, in steps of 50 GeV. At the generator level, we impose cuts on the $p_T$ of the jets (20 GeV) and of the leptons (20 GeV).
Additionally, we require jets and leptons to have $|\eta_j|<4.5$ and $|\eta_\ell|<2.5$.  With these baseline cuts, the $t\bar t$ background has a cross section at NLO of 1 nb, and the $W(Z)~+$ jets cross section is 220 nb (70 nb).\footnote{We have included in the $t\bar t+$ jets and in the $W(Z)$+jets cross sections a K-factor of 1.5~\cite{Campbell:2010ff} and 1.2~\cite{Campbell:2003hd}, respectively.} The $t\bar t Z$, $t\bar t W$ and 4 top background cross sections are much smaller: 1 pb~\cite{Maltoni:2015ena}, 770 fb~\cite{Campbell:2012dh}, and 12 fb~\cite{Bevilacqua:2012em}, respectively.  At the detector level, we have assumed a $b$-tagging efficiency approximately flat in $p_T$ and given by $70\%$ for $|\eta|<1.2$ and $40\%$ for $1.2<|\eta|<2.5$. This is what we call ``standard $b$-tagging" in Sec.~\ref{sec:2t2b}.

\begin{figure}[t]
\begin{center}
\raisebox{1.cm}{\includegraphics[width=0.5 \textwidth]{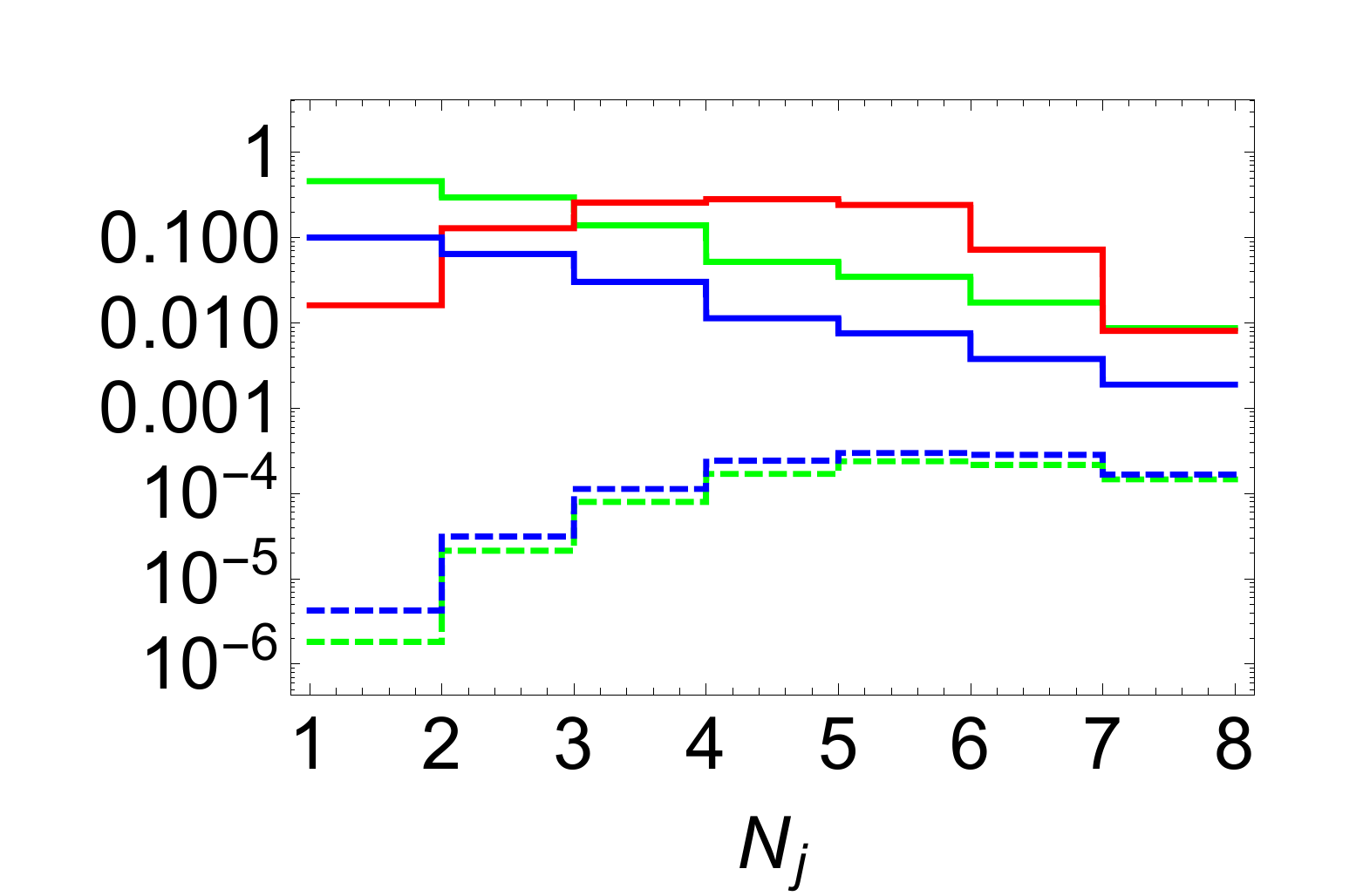}}
\includegraphics[width=0.45 \textwidth]{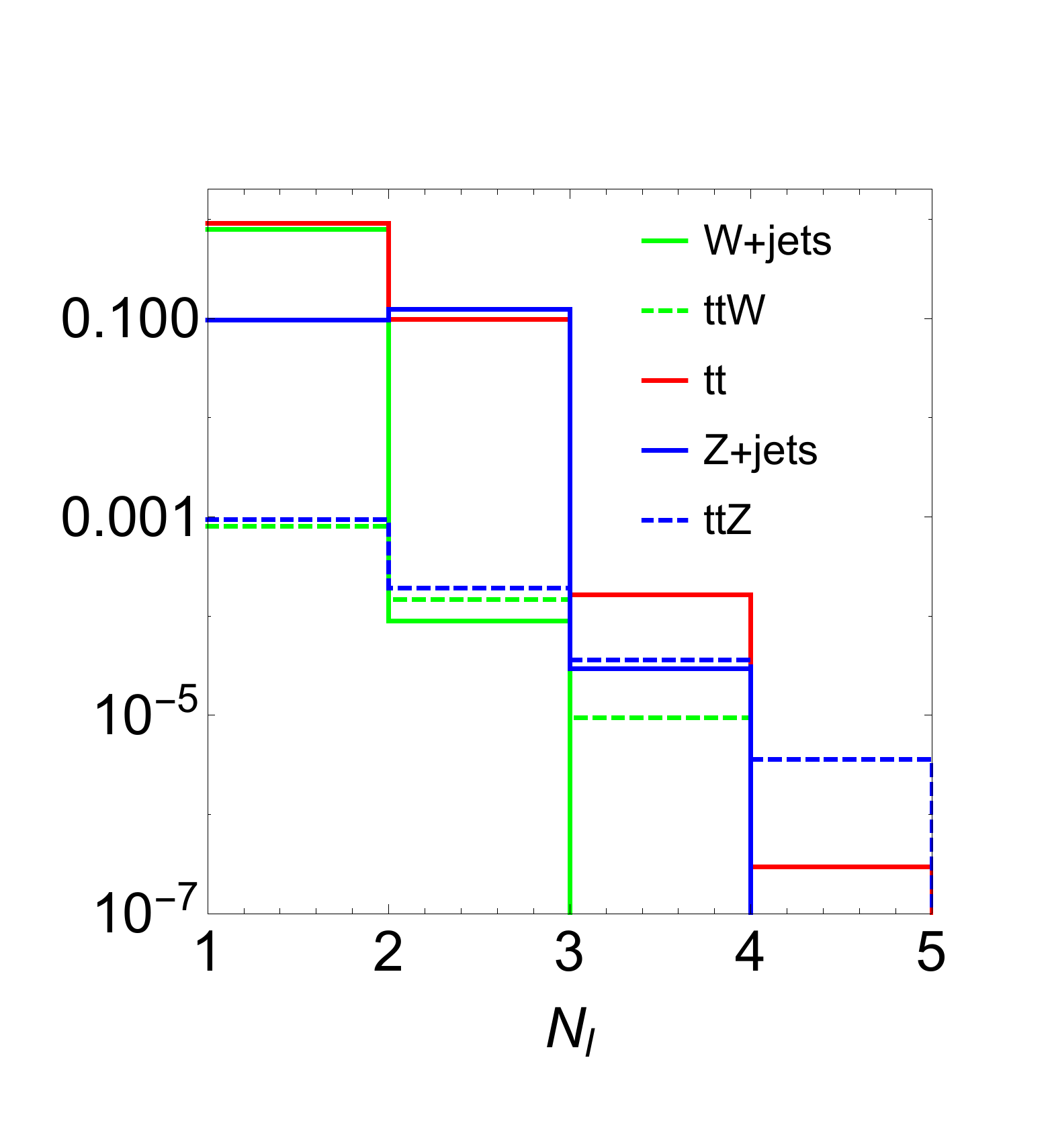}
\end{center}
\caption{Background distribution of the number of jets (left panel) and leptons (right panel) in each event. In the left panel, the fake rate we have used is $\epsilon_{200}=5\times 10^{-3}$ (see Sec.~\ref{sec:4t}). The events shown satisfy the requirement of at least one lepton and at least one $b$-tagged jet. In solid green we show the $W+$ jets background, in red the $t\bar t$ background, in solid blue the $Z+$ jets background, in dashed green the $t\bar{t}W$ background and in dashed blue the $t\bar{t}Z$ background. We normalize the distributions in such a way that the $t\bar t$ background integrates to 1, while the other backgrounds are normalized according to their cross section relative to $t\bar t$.}
\label{fig:Nj}
\end{figure}

In all the signal final states we study, the $t\bar t$+jets background is dominant.  This is because, for the $2b2t$ and $4t$ single lepton (Analysis (a.) below) signatures, we rely on high $H_T$ and a large multiplicity of jets (as well as at least one lepton and at least one $b$-tag). In the left panel of Fig.~\ref{fig:Nj}, we show the distribution for the number of jets in each event of the $t\bar t$+jets and $W(Z)$+jets backgrounds, after demanding at least one lepton and at least one $b$-tagged jet.  The total cross section of $t\bar t$ after these requirements is 150 pb. We scale the distributions in such a way that the $t\bar t$ background (in red) is normalized to 1 and the $W(Z)$+jets backgrounds (in green and blue, respectively) are rescaled according to their cross section relative to $t\bar t$.
As we can see from the figure, the $W$+jets background is rather sub-leading, as long as we require at least three jets. This is similar to what has been shown in Ref.~\cite{Lisanti:2011tm}, for the case of a one lepton and many jets signature (with no $b$-tag requirement).  In the figure for completeness we also show the results for the $t\bar t Z$ and $t\bar t W$ (in dashed blue and dashed green, respectively) even though they are negligible. 
 
The $t\bar t$+jets background  is also the dominant background for the $4t$, multi-lepton (Analysis (b.) below) signature, even though the exact estimation of the several backgrounds, in particular of the $t\bar t$+jets and $W(Z)$+jets, relies on the particular jet fake lepton rates adopted. In the right panel of Fig.~\ref{fig:Nj}, we show the distribution for the number of leptons in each event, after asking for at least one $b$-tagged jet and at least one lepton. The fake rate we have used is $\epsilon_{200}=5\times 10^{-3}$ (see Sec.~\ref{sec:4t} for details). As we can see from the figure, both the $W+$ jets and $Z+$ jets backgrounds are small, as long as we require at least three leptons. Asking for at least two $b$-tagged jets, as we do in our analysis of Sec.~\ref{sec:4t}, will further deplete the $W,Z+$ jets backgrounds, as compared to the $t\bar t$ background.  In the figure, we also show the results for the $t\bar t Z$ (dashed blue) and $t\bar t W$ backgrounds (dashed green). As can be seen, contrary to what was found for Analysis (a.), they represent a non-negligible background. Nevertheless, $t\bar t$ remains by far the dominant background. 
 
The invariant mass of the top quark pairs can be a relevant quantity for both the $2b2t$ and $2t$ case.  Since the top pairs in both these signal topologies arise from the decay of the heavy Higgs bosons, a different line shape in the invariant mass $m_{t\bar{t}}$ could, possibly, be a discriminant from the $t\bar{t}$ +jets SM background, for which a continuum distribution is expected. When analyzing both these signatures, we compute the invariant mass of the top quark pairs by first reconstructing the 4-momenta of the top quarks using the algorithm detailed in Appendix A and B of Ref.~\cite{Gresham:2011dg}, and summarized in the Appendix of this paper.

We now turn to analyzing each of the signatures in detail.

%%%%%%%%%%%%%%%%%%%%%%%%%%%%%%%%%%%%%%%%%%%%%%
\subsection{A Four Top Signature}\label{sec:4t}
%%%%%%%%%%%%%%%%%%%%%%%%%%%%%%%%%%%%%%%%%%%%%%

We begin by considering the $pp \rightarrow t \bar t H/A \rightarrow t \bar t t \bar t$ signature. 

\bigskip

As discussed above, the main SM background is given by $t\bar t$ + jets.  Because this background is large, and the signal cross-section is rather limited, as shown in Fig.~\ref{fig:BRttandSigma}, we will rely on either high jet multiplicities or multi-leptons to extract the signal over the background. 
This suggests two possible analysis strategies -- one analysis requiring at least one isolated lepton (either electron or muon) and at least six high $p_T$ jets (one of which must be $b$-tagged), and a second analysis requiring at least 3 leptons and 3 high $p_T$ jets (one of which is $b$-tagged). 
We require that the leptons have $p_T^\ell>10$ GeV,  $|\eta_\ell|<2.5$ and the jets (including $b$-jets) have $p_T^j>20$ GeV, $|\eta_j|<4.5$.  

\begin{figure}[tb]
\begin{center}
\begin{tabular}{ccc}
\includegraphics[width=0.45 \textwidth]{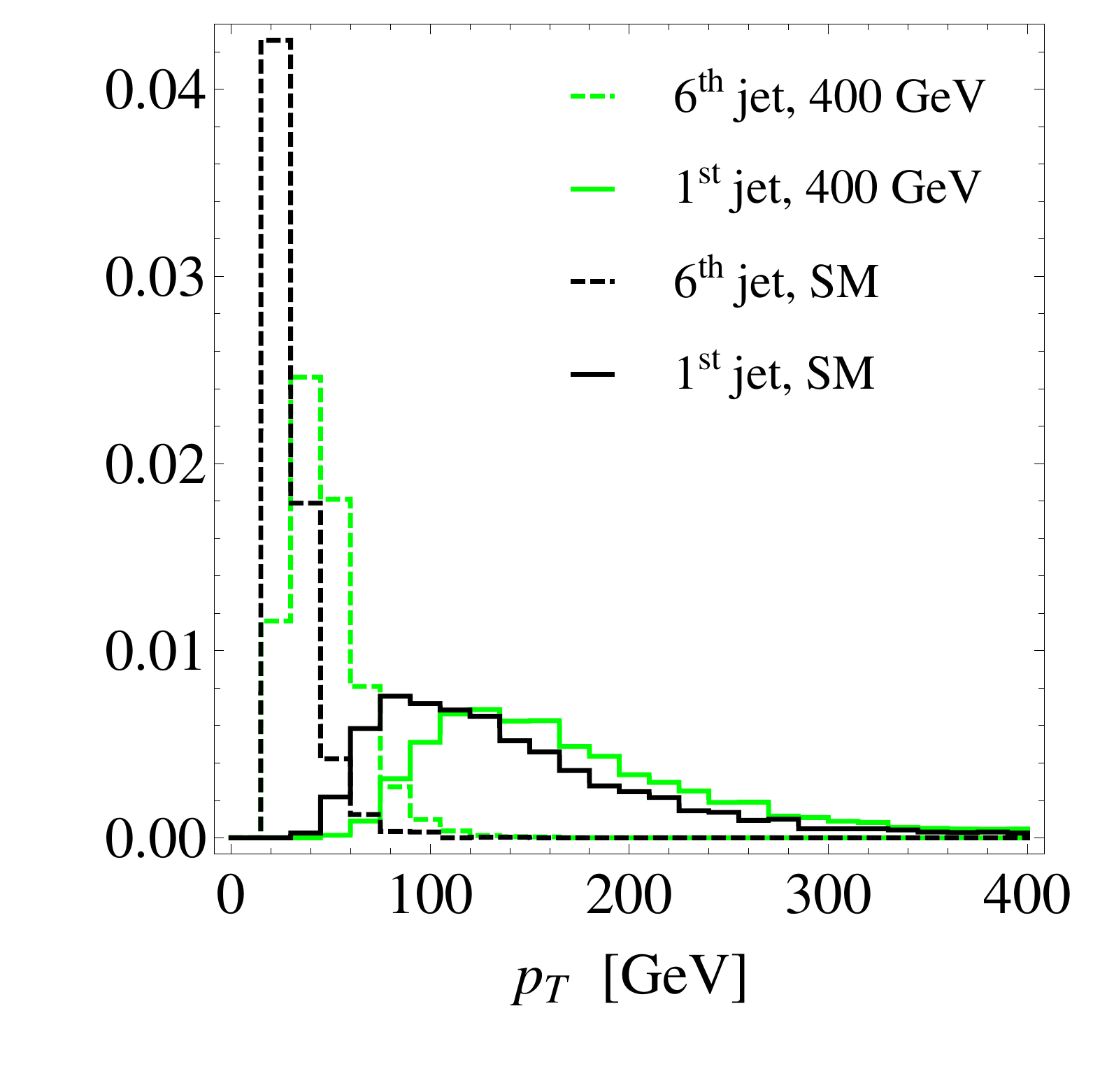}
& \hspace{4mm} &
\includegraphics[width=0.45\textwidth]{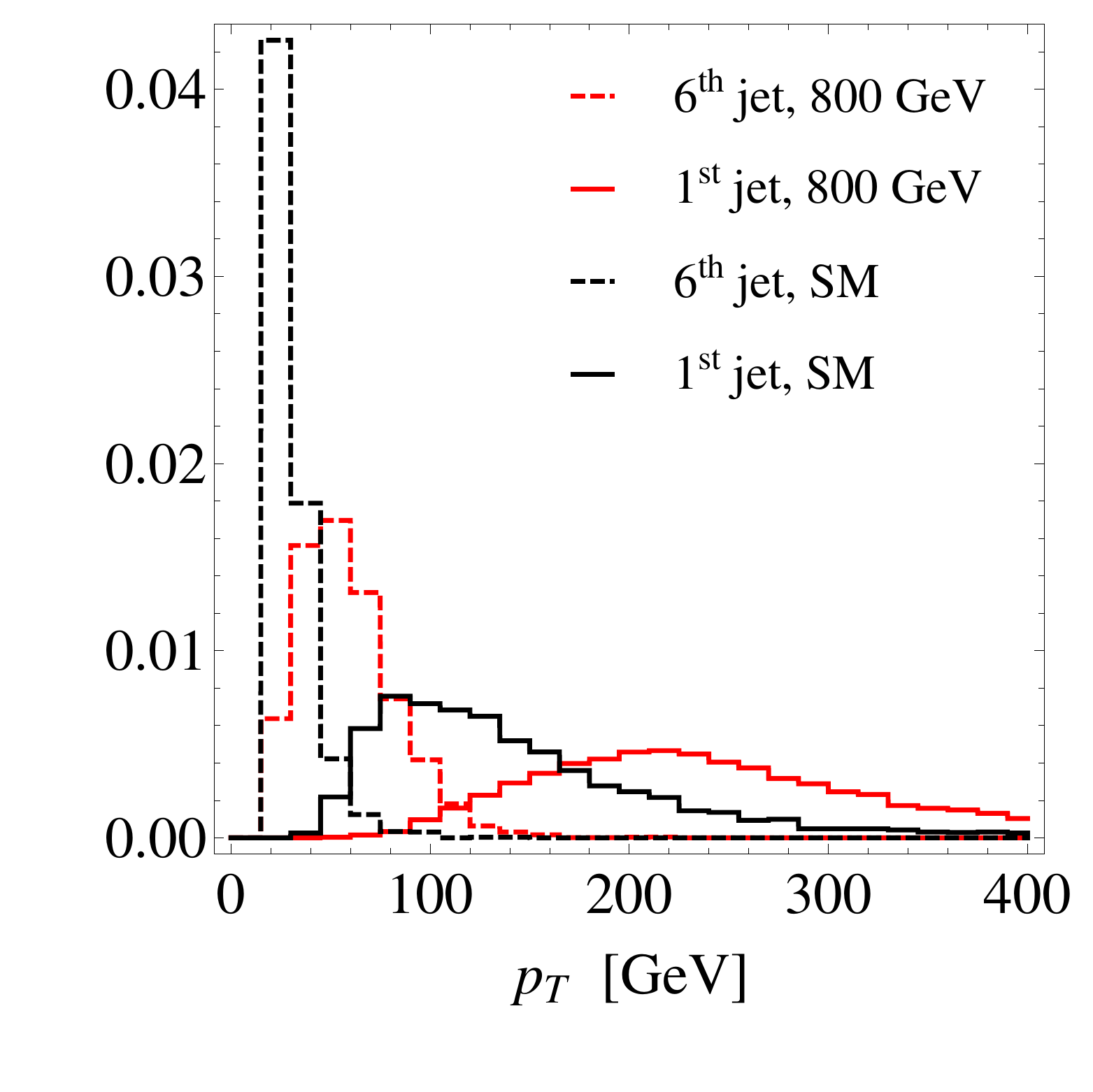}\\
\end{tabular}
\includegraphics[width=0.45 \textwidth]{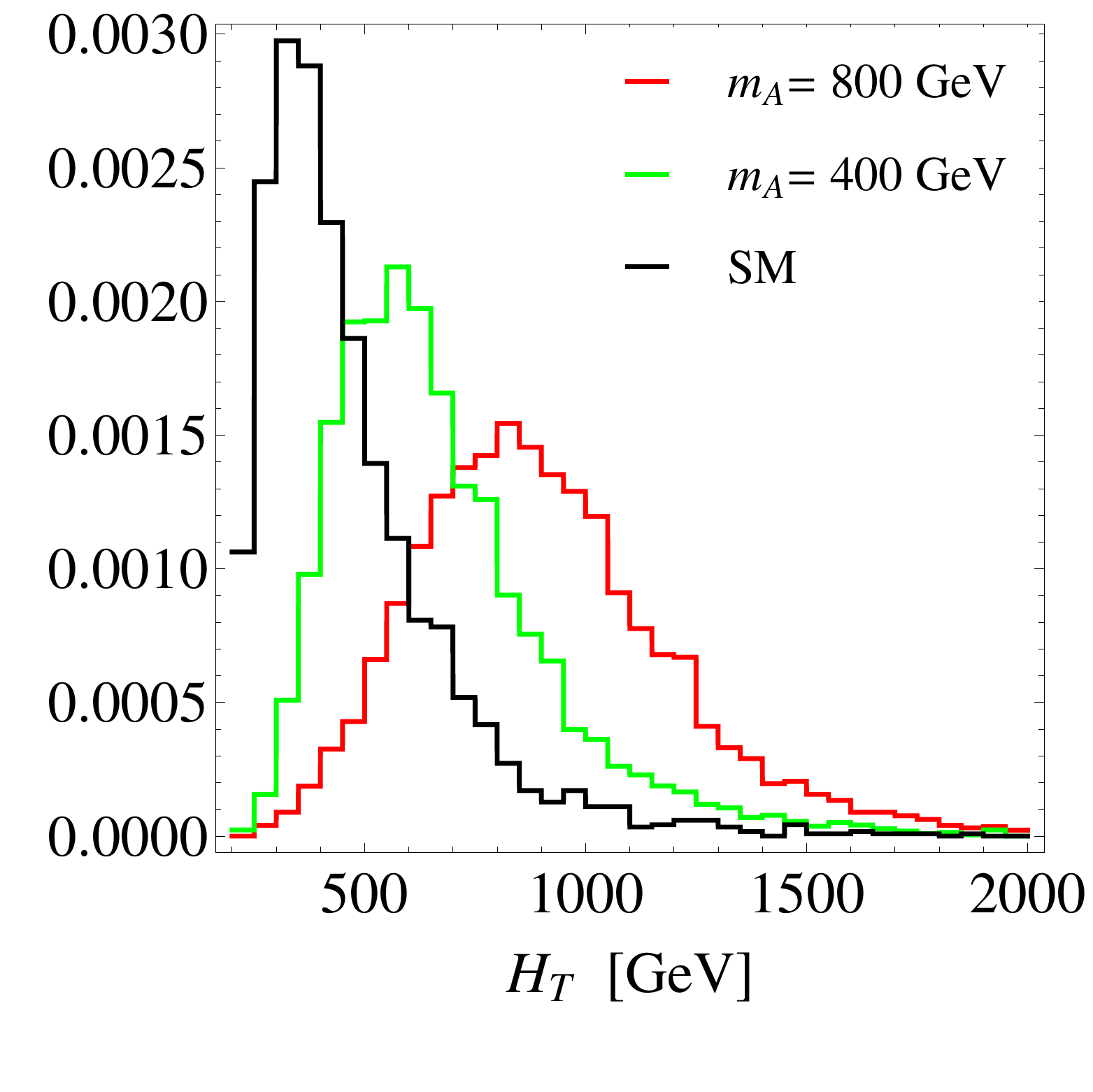}
\end{center}
\caption{ Upper plots: $p_T$ distribution of the leading jet (solid lines) and of the sixth jet (dashed lines) from a signal of a heavy Higgs with $m_A=400$ GeV (left plot) and with $m_A=800$ GeV (right plot). The distributions in black are the corresponding ones for the SM $t\bar t$ background. Lower plot: $H_T$ distribution for the signal ($m_A=400$ GeV in green, $m_A=800$ GeV in red) and for the background (in black). The distributions are normalized to one.}
\label{fig:pTjDistr}
\end{figure}

{\bf{Analysis (a.) -- Single Lepton.}}
We now discuss the single lepton plus high multiplicity jets analysis in detail. Similar searches have been performed at the 8 TeV LHC, constraining strongly produced gluinos in R-parity violating SUSY models \cite{CMS:2013qua}, as well as pair produced sgluons \cite{Aad:2015kqa}. 
In Fig.~\ref{fig:pTjDistr}, we show in green the $p_T$ distributions of the leading jet (solid lines) and of the sixth jet (dashed lines) from signal events of a 400 GeV heavy Higgs (upper left panel) and of a 800 GeV heavy Higgs (upper right panel). The corresponding distributions for the background are overlaid in black.  We use only events that pass our baseline cuts. All distributions are normalized to 1. As observed in the left panel, even a relatively light Higgs boson (400 GeV) produces a more boosted spectrum for the first six jets compared to the SM $t\bar t$ background. This feature is more evident for heavier Higgs bosons, as shown in the right panel of the figure.  In the lower panel of Fig.~\ref{fig:pTjDistr}, we show the $H_T$ distribution for the signal ($m_A=400$ GeV in green, $m_A=800$ GeV in red) and for the background (in black), where we have defined $H_T$ as the scalar sum of the $p_T$ of all jets in the event, including $b$-jets: $H_T=\sum p_T^j$.  As expected, the signal has higher $H_T$, especially for larger Higgs boson masses. 

Based on these considerations, we optimize our single lepton analysis as follows.
We generate signal events for benchmarks with heavy Higgs bosons with a mass in the range $m_A\in[400,1000]$ GeV, in steps of 50 GeV.  The variables that feed into our optimization are the $p_T$ of the leading jet ($p_{T1}$), 2nd, 3rd and 4th jet ($p_{T2}$), fifth and sixth jet ($p_{T3}$), the $p_T$ of the leading lepton ($p_{T\ell}$), $H_T$ and the number of $b$-jets ($N_b$).~\footnote{We have checked that replacing the $H_T$ variable with $m_{\rm{eff}}=\sum p_T^j+\sum p_T^\ell$ in our optimization does not lead to stronger constraints. }
We perform a 6-dimensional scan to find the cuts on these variables that maximizes the significance $S/\sqrt B$ at each value of $m_A$, while simultaneously requiring $S/B>1$ \% and 5\%. These two numbers for $S/B$ were chosen as proxies to demonstrate how well the LHC experiments can constrain the model parameter space given a particular handle on experimental and theoretical systematic uncertainties.   In particular, we build a 6-dimensional grid with $p_{T1}$ in the range [50-300] GeV with a step size of 50 GeV, $p_{T2,3}$ in the range [20-100] GeV with a step size of 20 GeV, $p_{T\ell}=(20,30)$ GeV, $H_T$ in the range [100-1000] GeV with a step size of 100 GeV, and finally $N_b=1,2,3$.  We have checked that increasing the maximum in the range of the $p_T$ of the jets and leptons
does not improve the reach of our proposed search.

\begin{figure}[tb]
\begin{center}
\includegraphics[width=0.49 \textwidth]{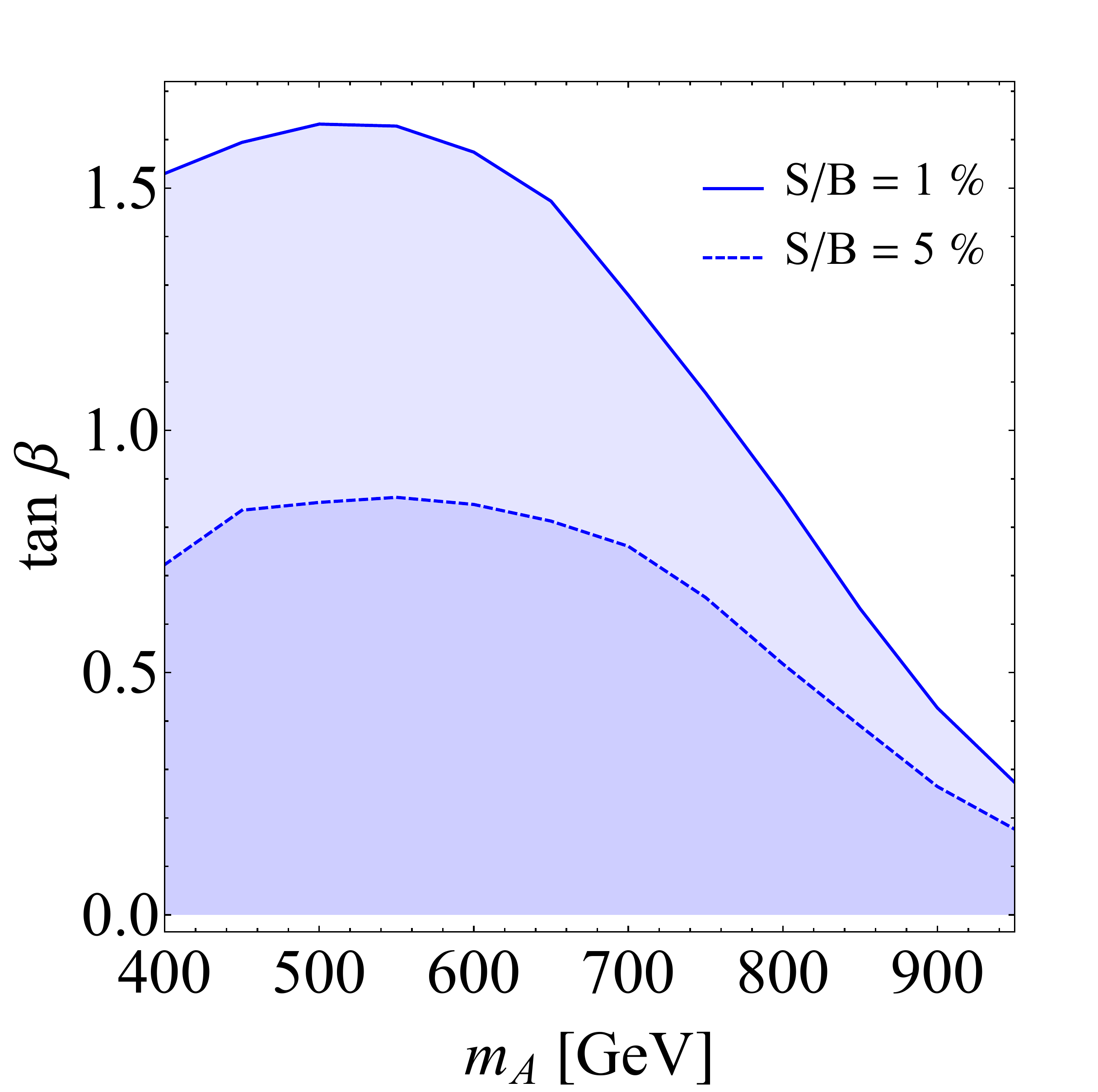}\; \;
\includegraphics[trim=1cm .2cm 1cm 1cm, clip=trip, width=0.39 \textwidth]{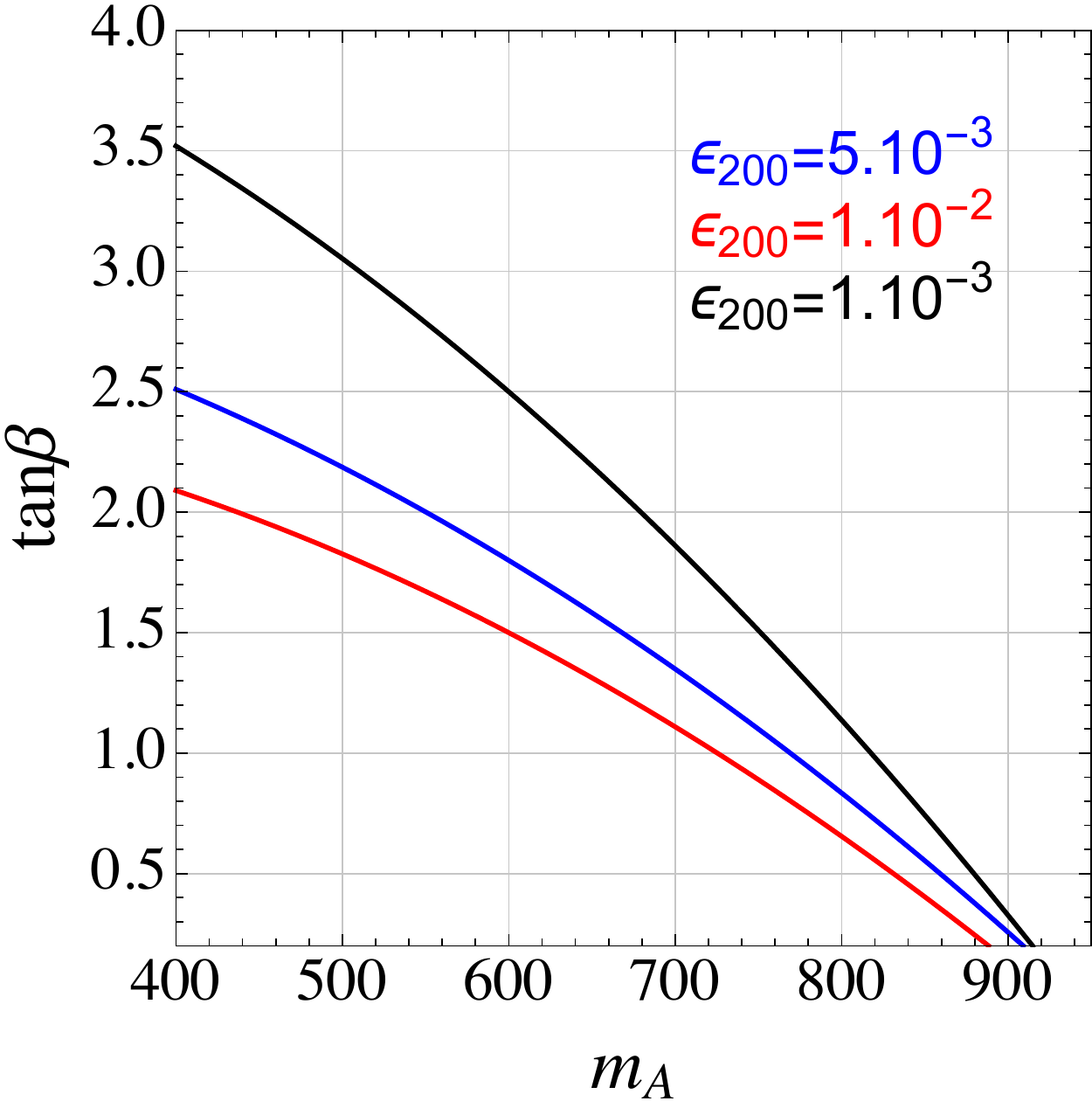}
\end{center}
\caption{Left: the $4t$ exclusion limit obtained using the single lepton analysis in the $m_A-\tan\beta$ plane, demanding $S/B=$ 1\% (solid blue boundary) or 5\% (dashed blue boundary). Right: the 95\% $4t$ exclusion limit obtained using the three lepton analysis  in the $m_A-\tan\beta$ plane. The three lines represent three different choices of fake rates:  $\epsilon_{200}=\{\epsilon_1,\epsilon_2,\epsilon_3\}=\{5\times 10^{-3},\;  10^{-2},\;  10^{-3}\}$.}
\label{fig:mAT4t_1l}
\end{figure}

To show the impact of these cuts, in Table~\ref{tab:forpol} (a) we show the cut flow table corresponding to our optimal cuts for the benchmark scenario with $m_A=400$ GeV and $\tan\beta=1.5$. 
A sizable improvement in the $S/B$ ratio arises once we demand at least 3 $b$-jets. In the left panel of Fig.~\ref{fig:mAT4t_1l}, we show the reach of our proposed search: values of $\tan\beta$ as large as $\sim 1.5$ can be probed for heavy scalar masses up to $\sim 600$ GeV. The main issue that the ATLAS and CMS collaborations will have to overcome is systematic uncertainties.  To show this effect, in the left panel of Fig.~\ref{fig:mAT4t_1l} we compare the expected exclusion after requiring $S/B$= 1\% (solid blue boundary) and 5\% (dashed blue boundary).   As we can see in the figure, the bound will depend strongly on the degree to which systematic uncertainties can be reduced in the High Luminosity runs.

\begin{table}
\begin{minipage}[b]{.45\textwidth}
  \centering
  \begin{tabular}{ | l || l | l|l|l|}
    \hline\hline
    Cuts & 400 GeV &  $t\bar{t}+$ jets  & $S/B$ & $S/\sqrt B$\\ 
       1$l$+$b$-jets  & ($21.3$ fb) & ($1$ ab) & &\\\hline
    Baseline   &    19600      &   $5.0\times 10^7$    & $3.9\times 10^{-4}$        &  2.8       \\ \hline
    $p_{T1}>200$ & 6200 & $9.5\times 10^6$ & $6.6\times 10^{-4}$ & 2.0 \\ \hline
      $p_{T2}>60$ & 5200 & $5.6\times 10^6$ & $9.2\times 10^{-4}$  & 2.2 \\ \hline
        $p_{T3}>60$ & 1900 & $1.2\times 10^6$ & $1.6\times 10^{-3}$ &1.8 \\ \hline
        $H_T>800$ & 1700 & $1.0\times 10^6$ & $1.7\times 10^{-3}$ & 1.7 \\\hline
          $N_b\geq3$  & 400 & $3.9\times 10^4$ & $1.0\times 10^{-2}$  & 2.0\\ \hline
    \hline
  \end{tabular}
\end{minipage}\qquad
\begin{minipage}[b]{.5\textwidth}
  \centering
  \begin{tabular}{|c||c|c|c|c|}
 \hline
 \hline
 & $400$ GeV &  $t\bar{t}+$ jets & $S/B$ &  $S/\sqrt B$   \\
 3$l$+$b$-jets & ($21.3$ fb)  & ($1$ ab) && \\
 \hline
$\epsilon_1$ &240&1850 &  0.13 & 5.6 \\ 
 $\epsilon_2$  &240 &  3870 & 0.06 & 3.8 \\
 $\epsilon_3$ & 240& 480 & 0.5 & 10.9 \\
 \hline
  \end{tabular}
\end{minipage}\vspace{0.5cm}
(a)  \hspace{3in} (b)
\caption{Cut flow tables for the number of events  at  LHC14 with 3000 fb$^{-1}$ data arising from a representative scenario for the $4t$ signature: $m_H=400$ GeV and $\tan\beta=1.5$. In the tables, we sum the contributions of the scalar $H$ and the pseudoscalar $A$.  {\bf{Left}}: Analysis (a.) One lepton + multi-$b$-tag signature. {\bf{Right}}: Analysis (b.) Three leptons and two $b$-tag signature.  In the latter case three different fake rates are used for the background: $\epsilon_{200}=\{\epsilon_1,\epsilon_2,\epsilon_3\}=\{5\times 10^{-3}, \; 10^{-2},\;  10^{-3}\}$. }  \label{tab:forpol}
\end{table}

\bigskip

{\bf{Analysis (b.)--Multi-Lepton.}}
Next we consider the constraints on a heavy Higgs boson when requiring at least three leptons in addition to the jets (including $b$-tags).  Similar searches have been performed at the 8 TeV LHC, such as searches for strongly produced gluinos \cite{CMS:2014dpa}. As shown in the right panel of Fig.~\ref{fig:Nj}, the main backgrounds for this signature are $t\bar{t}Z$ and $t\bar t+$ jets, where at least one jet fakes a lepton.  ATLAS and CMS collaborations use a data-driven  approach  to  estimate  lepton  fakes
in  their  multilepton  analyses~\cite{CMS:2013ida,ATLAS3l13}. Since  we  do not have access to the resources needed for data-driven
estimates, we adopt the approach proposed in Ref.~\cite{Curtin:2013zua}. This method exploits the relationship between the kinematics
of a fake lepton and that of the heavy flavor jet that produces it. In particular, we apply to each heavy flavor jet a probability of generating a fake lepton ($\epsilon_{200}$), assumed to be a function of the jet $p_T$, and a transfer
function ($\mathcal{T}_{j\to \ell}(\alpha)$), which represents a normalized probability distribution for the fraction of the jet momentum ($\bar p^j$)
that is inherited by the fake lepton ($\bar p^\ell$):

\begin{eqnarray}
\epsilon_{j\to \ell}(p_T^j) &=&\,\, \epsilon_{200}\left[1-(1-r_{10})\frac{200-p_T^j/\mathrm{GeV}}{200-10}\right]\,, \\
\mathcal{T}_{j\to \ell}(\alpha) &=&\, \left(\frac{\sqrt{2\pi}\sigma}{2}\right)^{-1} \left[\mathrm{erf}\left(\frac{1-\mu}{\sqrt{2}\sigma}\right)+\mathrm{erf}\left(\frac{\mu}{\sqrt{2}\sigma}\right)\right]^{-1}e^{-\frac{(\alpha-\mu)^2}{2\sigma^2}},
\end{eqnarray}
where $\alpha\equiv 1-\bar p^\ell / \bar p^j\,$ is the fraction of the heavy flavor jet momentum that is not transferred to the fake lepton, and $r_{10}$ parametrizes the dependence of the fake rate on the transverse momentum of the jet.

For simplicity, we consider an efficiency independent of the jet $p_T^j$ ($r_{10}=1$). We further set $\mu=0.5$, 
 based on
the expectation of roughly equal splitting of the momentum between the fake lepton and the
neutrino produced in heavy 
flavor decays. Finally, we consider values of $\sigma$ and $\epsilon_{200}$ such that we reproduce the 8 TeV CMS three lepton, $\geq 2$ $b$-jet signal regions in Ref~\cite{CMS:2013ida}. We find that $\sigma=0.1$ and $\epsilon_{200}=5\times 10^{-3}$ gives a good fit to the data.

Applying these parameters for fake leptons, we compute the $t\bar t+$ jets and $W(Z)$+jets background cross sections at 14 TeV LHC, requiring at least 3 leptons, 2 $b$-jets and 3 ordinary jets (including $b$-jets).  Our results for the number of events at the 14 TeV LHC with 3000 fb$^{-1}$ data are reported in Table \ref{tab:forpol} (b) (row $\epsilon_1$ in the table). To assess the impact of a different fake rate at LHC14, we also report our results for two additional fake rates: $\epsilon_2=10^{-2},\,\epsilon_3=10^{-3}$. In the table, we do not report the values of the  $W(Z)$+jets and $t\bar{t}W(Z)$ backgrounds, since these are sub-leading and at least one order of magnitude smaller than the $t\bar t$ background, for every choice of the fake rate $\epsilon_{200}$ (see also the right panel of Fig.~\ref{fig:Nj}).
In the right panel of  Fig.~\ref{fig:mAT4t_1l}, we present the reach in the $m_A-\tan\beta$ plane for these three different choices for the fake rates $\epsilon_i$.  The reach of this proposed search is more robust than the reach of the single lepton search, even though it greatly depends on the lepton fake rates ATLAS and CMS will be able to achieve at the High Luminosity stage. In particular, for our three choices of fake rates $(\epsilon_1,\epsilon_2,\epsilon_3)$, we can probe heavy Higgs bosons up to 770, 720, 820 GeV, respectively, for $\tan\beta=1$.

Quantitatively, the main difference between the 3 lepton and 1 lepton analysis, as can be seen comparing the two tables in Table~\ref{tab:forpol}, is that $S/B$ is greatly improved in the 3 lepton analysis, highlighting the leading role that systematics on the $t\bar t$ background plays in limiting the reach of the 1 lepton search.  

In the next section, we will see that the $t\bar t$ background systematics will be even more challenging for designing a search for heavy Higgs bosons in the $2b2t$ final state.

%%%%%%%%%%%%%%%%%%%%%%%%%%%%%%%%%%%%%%%%%%%%%%%%%%%%%%%%%%%%
\subsection{A Two Top, Two Bottom Signature}\label{sec:2t2b}
%%%%%%%%%%%%%%%%%%%%%%%%%%%%%%%%%%%%%%%%%%%%%%%%%%%%%%%%%%%%%

We next turn to considering the $2b2t$ final state.  As can be seen from Fig.~\ref{fig:BRttandSigma},  compared to the $4t$ signature presented in the previous section, the signal cross section for $pp\to b\bar b H(A), H(A)\to t\bar t$ is significantly larger. At the same time, the signal is harder to disentangle from the $t\bar t$+jets background. This is due to a lower multiplicity of jets, the low $p_T$  
of  many of the $b$-jets produced in association with the Higgs boson, and the challenge of $b$-tagging such jets which are additionally produced at relatively high pseudorapidity.  

\begin{figure}[phtb]
\begin{center}
\includegraphics[width=0.9 \textwidth]{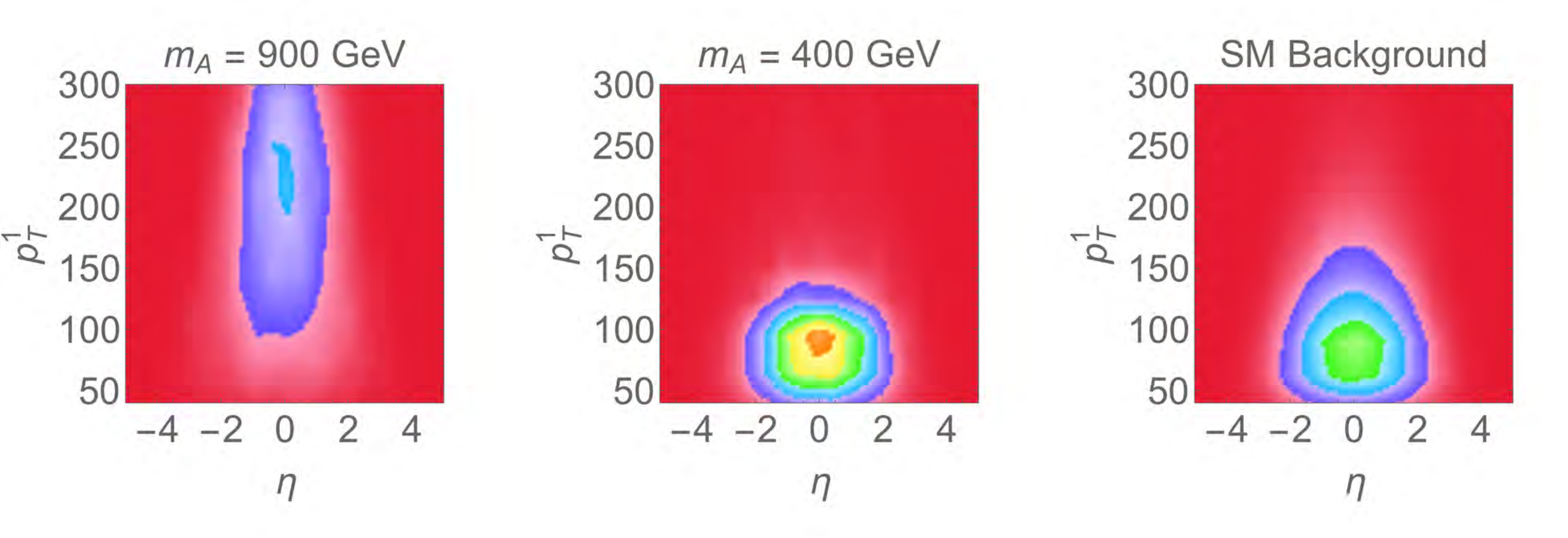} \includegraphics[trim=0.4cm .1cm 0.1cm .1cm, clip=trip, width=0.05 \textwidth]{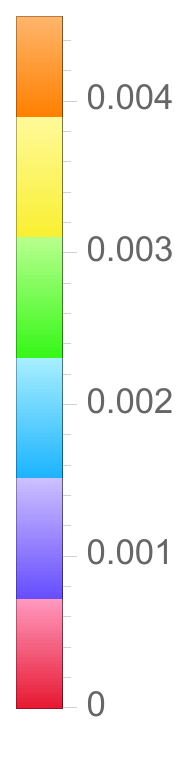}
\includegraphics[width=0.9 \textwidth]{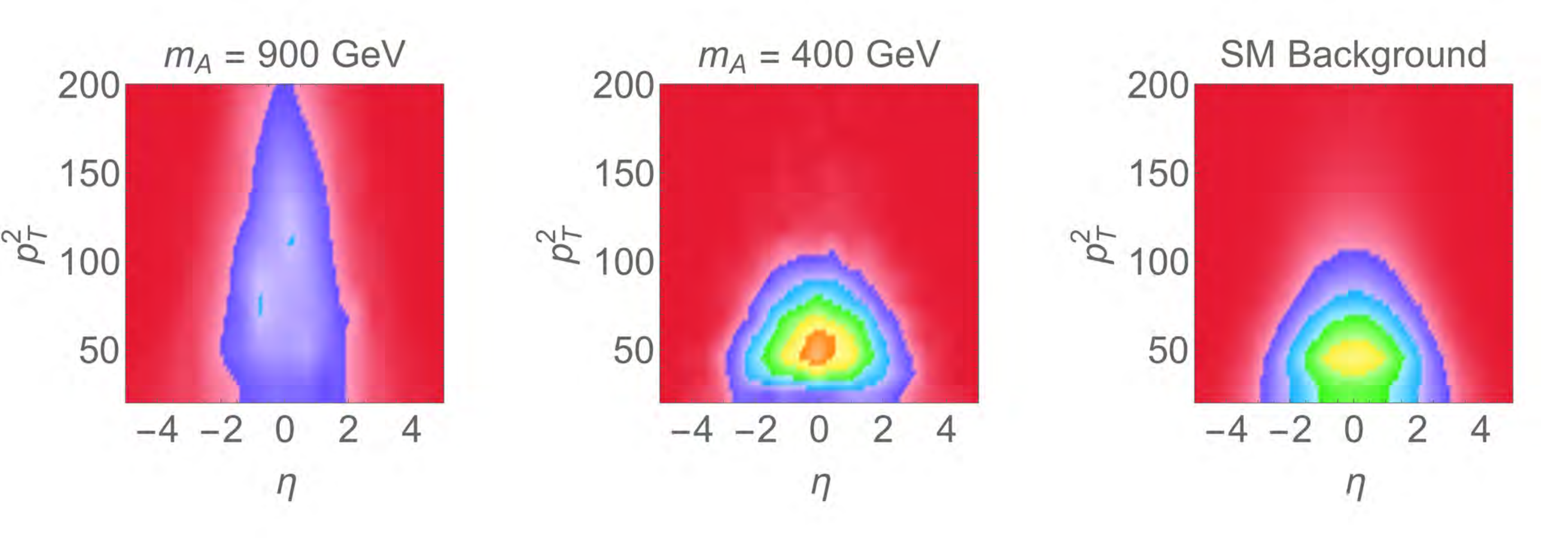} \includegraphics[trim=0.4cm .1cm 0.1cm .1cm, clip=trip, width=0.05 \textwidth]{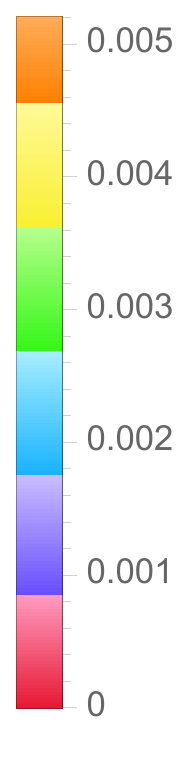}
\includegraphics[width=0.9 \textwidth]{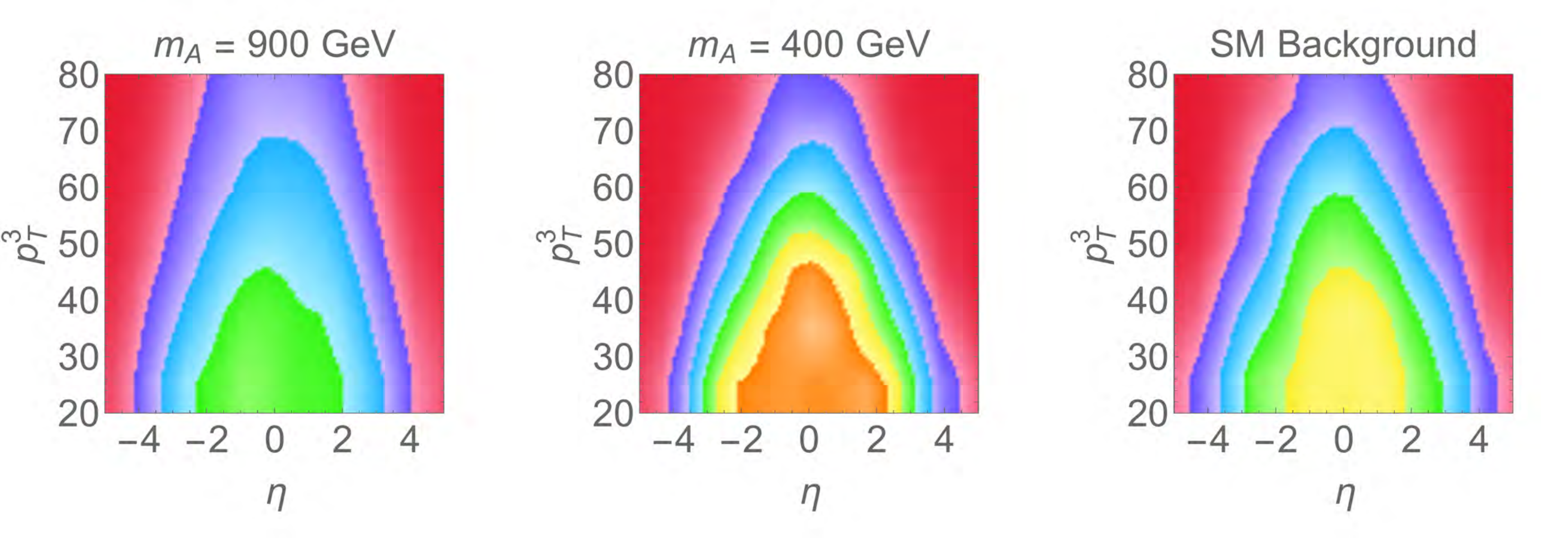} \includegraphics[trim=0.4cm .1cm 0.1cm .1cm, clip=trip, width=0.05 \textwidth]{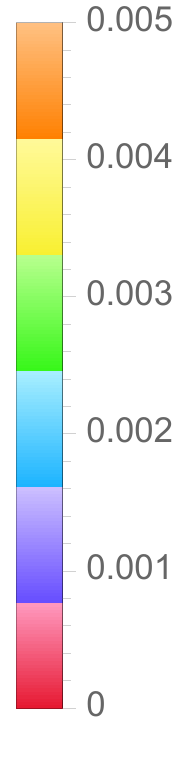}
\includegraphics[width=0.9 \textwidth]{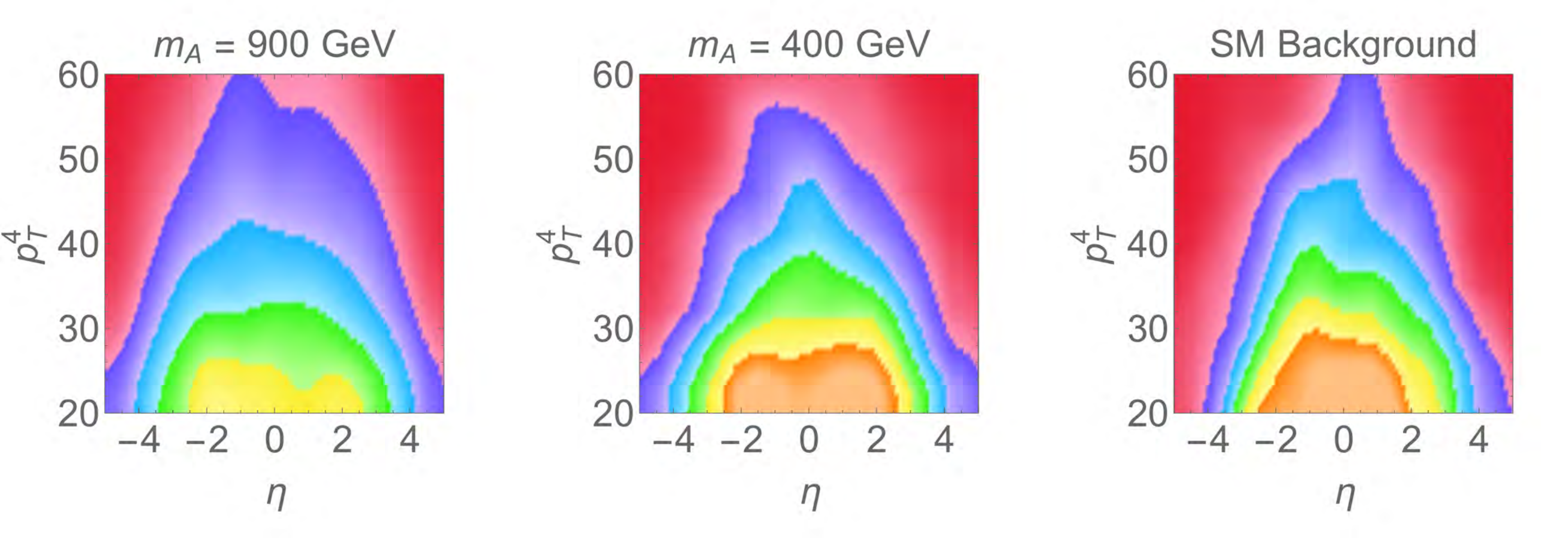} \includegraphics[trim=0.4cm .1cm 0.1cm .1cm, clip=trip, width=0.05 \textwidth]{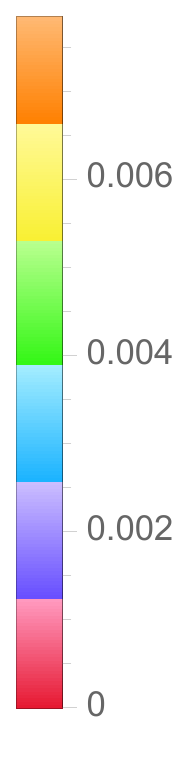}
\end{center}
\caption{$p_T$ versus $\eta$ distributions, as obtained from parton level events for the four hardest $b$ quarks in two representative signal scenarios ($m_H=900$ GeV, left column, $m_H$=400 GeV, middle column), as well as in the SM $t\bar{t}$ background (right column).}
\label{fig:bquarksttbb}
\end{figure}

These features can be seen in Fig.~\ref{fig:bquarksttbb}, where the $p_T$ versus $\eta$ distributions are shown, as obtained from parton level events for the four hardest $b$ quarks in two representative signal scenarios ($m_H=900$ GeV, left column, $m_H$=400 GeV, middle column) as well as in the SM $t\bar{t}$+jets background (right column). 
Typically the two $b$ quarks with lowest $p_T$ $(p_T^3,\,p_T^4)$ are the $b$ quarks produced in association with the Higgs boson.  As shown in Fig.~\ref{fig:bquarksttbb}, these two $b$ quarks have low $p_T$, and reach relatively sizable values of the pseudorapidity. In fact, we find that, almost independent of the $H/A$ mass, the signal efficiency at the parton level after the simple requirement that these two $b$-quarks have $p_T$ larger than 20 GeV and $|\eta|<2.5$ is $\sim 15\%$, emphasizing the challenge of pursuing this signature.  As shown in the top two panels of Fig.~\ref{fig:bquarksttbb}, the two hardest $b$ quarks can instead represent a better discriminant between the signal and the background, at least for relatively heavy Higgs bosons.

\begin{figure}[t]
\begin{center}
\begin{tabular}{ccc}
\includegraphics[width=0.45 \textwidth]{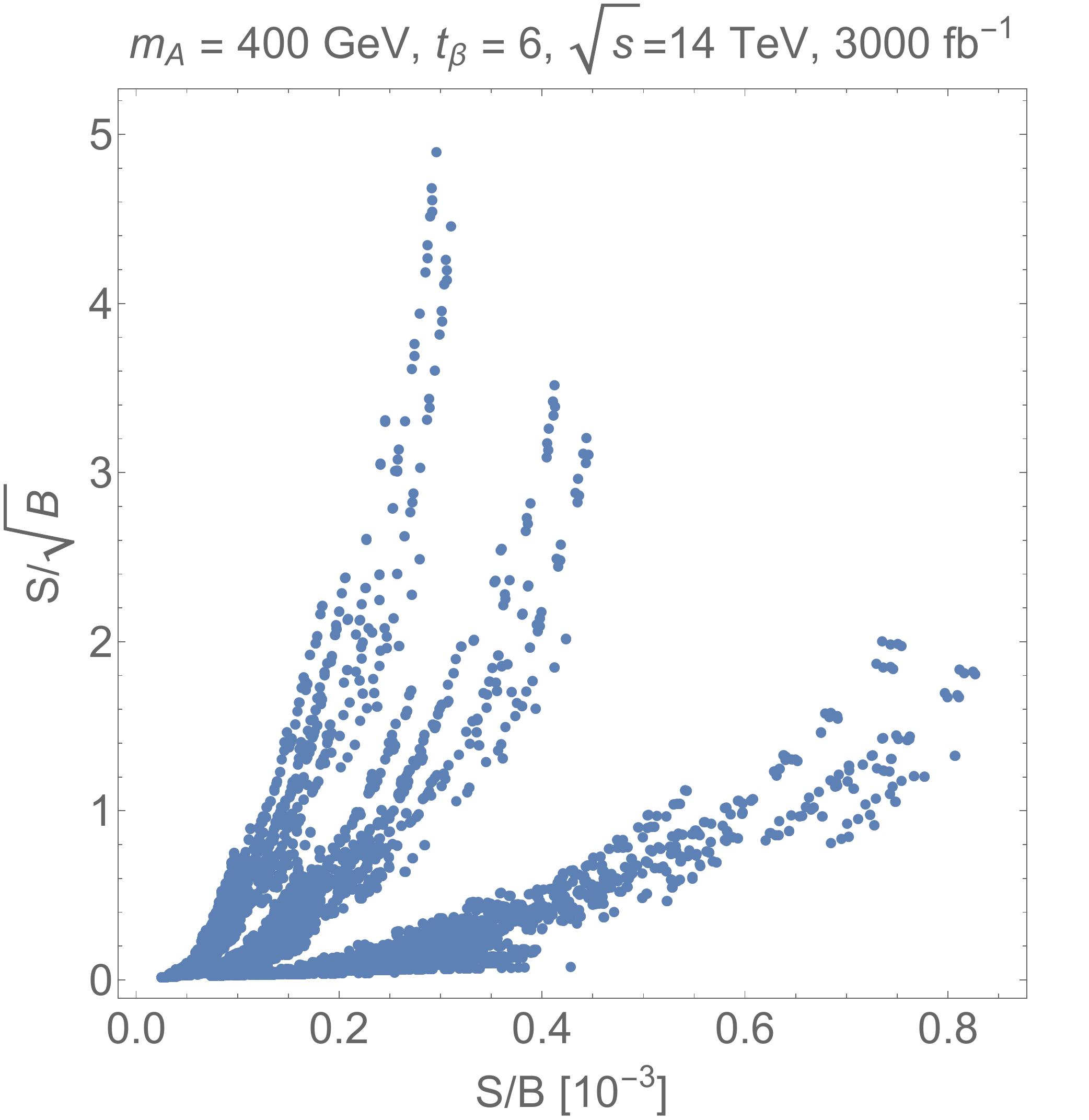}
& \hspace{4mm} &
\includegraphics[width=0.45\textwidth]{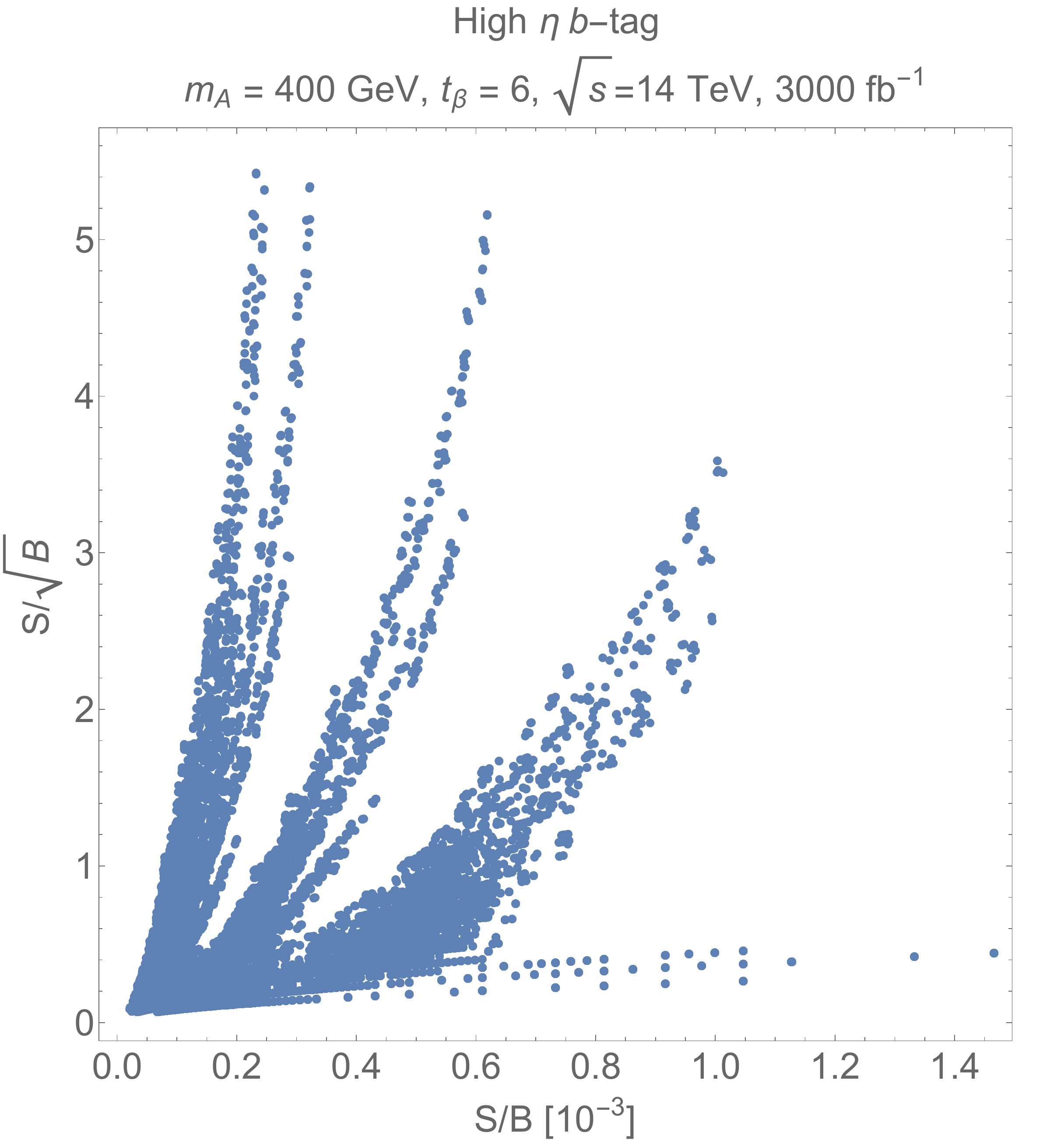}\\
\end{tabular}
\end{center}
\caption{$S/B$ versus $S/\sqrt{B}$ for the $2b2t$ signature for the benchmark scenario with $t_\beta = 6$ and $m_A =400$ GeV. Each point corresponds to an element of our scan,
varying the cuts on the $p_T$ of the jets ($p_{T1},\,p_{T2,3}$) and of the lepton ($p_{T\ell}$), as well as the number of $b$-tags and $H_T$ (see text for details of the scan).  In the left-hand panel, a standard $b$-tag is applied, while in the right-hand panel $b$-tagging allowing for higher values of $\eta$ (3.7) is applied. The three (four) branches correspond to three (four) $b$-tags in the left(right)-hand plot.}
\label{SB}
\end{figure}

\begin{figure}[t]
\begin{center}
\includegraphics[width=0.6 \textwidth]{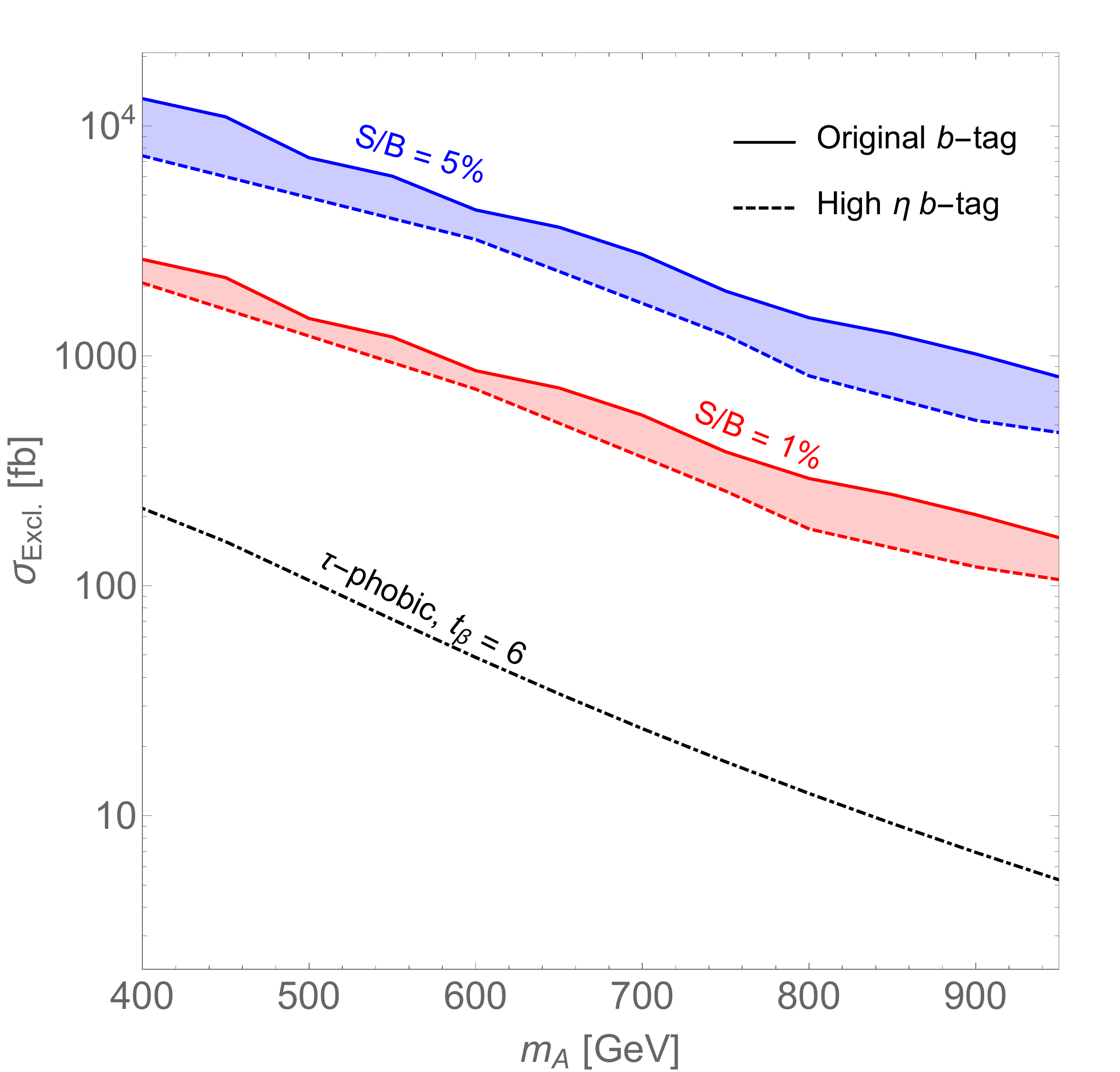}
\end{center}
\caption{Excluded cross-section as a function of $m_A$ with $S/B=1\%$ (red) and $=5\%$ (blue). The upper (solid) and lower (dotted) bounds of each band represent our results obtained with the standard and high-$\eta$ $b$-tagging, respectively.   The dot-dashed black curve shows the cross section for $pp\to b\bar b H/A,\,H/A\to t\bar t$ in the $\tau$-phobic scenario, having fixed $t_\beta = 6$.}
\label{2b2t_Exc}
\end{figure}

Similar to the $4t$ signature, we generate signal events for benchmarks with heavy Higgs bosons with a mass in the range $m_A\in[400,1000]$ GeV, in steps of 50 GeV.   
The variables that enter our optimization are the $p_T$ of the leading jet ($p_{T1}$), 2nd jet ($p_{T2}$), and 3rd and 4th jet ($p_{T3}$), the $p_T$ of the leading lepton ($p_{T\ell}$), $H_T$ and the number of $b$-jets ($N_b$). 
In particular, we build a 6-dimensional grid with $p_{T1}$ in the range [50-300] GeV with a step size of 50 GeV, $p_{T2,3}$ in the range [20-100] GeV with a step size of 20 GeV, $p_{T\ell}=(20,30)$ GeV, $H_T$ in the range [100-1000] GeV with a step size of 100 GeV and finally $N_b=1,2,3$.   We have checked that scanning over larger values for $p_{T1},\,p_{T2,3},\,p_{T\ell},\,H_T$, as well as including $N_b=4$, does not lead to a better bound on the excluded cross section (see Fig.~\ref{2b2t_Exc}). We show the $S/B$ versus $S/\sqrt{B}$ resulting from this scan in the left panel of Fig.~\ref{SB}  for $m_A$ = 400 GeV and $t_\beta$=6. This value of $\tan\beta$ has been chosen to maximize the $pp\to b\bar b H/A,\,H/A\to t\bar t$ cross section (see Fig.~\ref{fig:BRttandSigma}). The three branches shown in the figure represent the results of our scan, having fixed $N_b=1,2,3$, from top to bottom.  One can see that, while $S/\sqrt{B}$ can easily be sufficiently large ($\geq 2$), $S/B$ is, instead, typically much smaller than what might possibly be obtained theoretically or experimentally, where a systematic uncertainty of at least several percent is expected.  We also checked if reconstructing the $t \bar t$ resonance, using the top reconstruction algorithm outlined in  Appendix A, could improve the outlook; we found that the smearing from detector effects as well as combinatorics imply only a small improvement (see Appendix A for details). 

One might wonder if one could better utilize the presence of the forward and low $p_T$ $b$-jets from the initial state to discriminate signal from background.  To do this effectively would require increased $b$-tagging efficiency at higher rapidities and smaller $p_T$.  Motivated by the ATLAS Phase-II Upgrade Scoping Document~\cite{ATLASUpgrade}, we consider a $b$-tagging efficiency of $70\%$ for $|\eta|<2.5$ and $40\%$ for $2.5<|\eta|<3.7$.  We (very) optimistically assume that the $b$-tagging efficiency is independent of $p_T$, down to 20 GeV.   In the right panel of Fig.~\ref{SB}, we show the results of our scan with this improved $b$-tagging efficiency, including also $N_b=4$ in our scan (lowest branch in the figure).  Modifying the $b$-tagging efficiency does indeed improve $S/B$, and in particular, allows us to obtain a reasonably sizable signal for 4 $b$-jets in the events. These are the points shown with the largest $S/B$. However, as can be seen, this gain is not sufficient to guarantee values of $S/B$ at the percent level. For this reason, we do not expect that this search will be an efficient probe of the
2HDM $(m_A-\tan\beta)$ plane, at the 14 TeV LHC with 3000 fb$^{-1}$ of data.

More general models could produce a larger cross section for a resonance produced in association with $b$ quarks and decaying into tops, $pp\to b\bar b H/A,\,H/A\to t\bar t$. For this reason, in Fig.~\ref{2b2t_Exc}, we show the excluded cross-section with $S/B=1\%$ (red) and $=5\%$ (blue). The upper (solid) and lower (dotted) bounds of each band represent our results obtained with standard and high-$\eta$ $b$-tagging, respectively. 
For this plot, we chose cuts based on our scan so as to maximize $S/\sqrt B$, with the requirement $S/B\geq 1\%$ (in red),  $S/B\geq 5\%$ (in blue). For comparison, the dot-dashed black curve shows the cross section for $pp\to b\bar b H/A,\,H/A\to t\bar t$ in the $\tau$-phobic scenario, having fixed $t_\beta =6$, to maximize the cross section.  As can easily be seen, the large systematic uncertainties do not allow one to put meaningful constraints on the heavy Higgs in the $2b2t$ channel.

Two comments are in order: the $b$-tagging at higher values of $\eta$ allows us to probe cross sections up to a factor of two smaller than the standard $b$-tagging. In addition, in the optimistic scenario in which the LHC collaborations will be able to reduce systematics to the level of $S/B\sim 1\%$ in the High Luminosity runs, cross sections one order of magnitude larger than the cross sections of our $\tau$-phobic scenario can be probed.

%%%%%%%%%%%%%%%%%%%%%%%%%%%%%%%%%%%%%%%%%%%%%%%%%%%%%%%%%%%%
\subsection{Top Resonance Searches}\label{sec:2t}
%%%%%%%%%%%%%%%%%%%%%%%%%%%%%%%%%%%%%%%%%%%%%%%%%%%%%%%%%%%%%

The case of the $2t$ signature is fundamentally different in nature from the previous two signatures discussed. In particular, the interference between the SM QCD production and resonant $t \bar t$ production from $A,~H$ is fundamentally important. This was calculated first in Ref.~\cite{Dicus:1994bm} and has been revisited recently in Ref.~\cite{Craig:2015jba,Jung:2015gta}. This interference gives rise to a distinctive peak and dip structure that one may hope to be able to use to observe the presence of a heavy scalar or pseudoscalar resonance. Ref.~\cite{Jung:2015gta} utilizes the current 8 TeV $t\bar t$ searches to project LHC14 reach, having only the total deficit in $t\bar{t}$ production due to almost pure-dip structure from the interference. It has been shown that sensitivity may be obtained for deficits of order of 10's of fb for Higgs boson masses less than 1 TeV with 3000 fb$^{-1}$ of data. Ref.~\cite{Craig:2015jba}, instead, shows that, for the case where both a peak and a dip is present, detector effects will in all likelihood wash out this structure so as to make it very difficult to disentangle from the SM background. The analysis carried out in Ref.~\cite{Craig:2015jba} relied on a smearing function that was obtained by simulating SM (only) events both at the parton level and after showering and detector simulation; no signal events were utilized to obtain the smearing function. 

We refine this analysis by implementing for the first time in a Montecarlo tool,  {\tt MadGraph},  the scalar/pseudoscalar interference with the QCD background.  This allows us to take into account efficiencies for the signal and background when fully interfered first at parton level, then showered.  Detector effects were applied to the combined signal and background events.  This procedure when applied to the fully interfered sample was numerically very intensive, given the small size of the interference in comparison to the background.   It does give us confidence, however, that we have fully taken into account the detector and $t \bar t$ reconstruction effects on the fully hadronized sample. 

\begin{figure}[tb]
\begin{center}
\includegraphics[width=0.75 \textwidth]{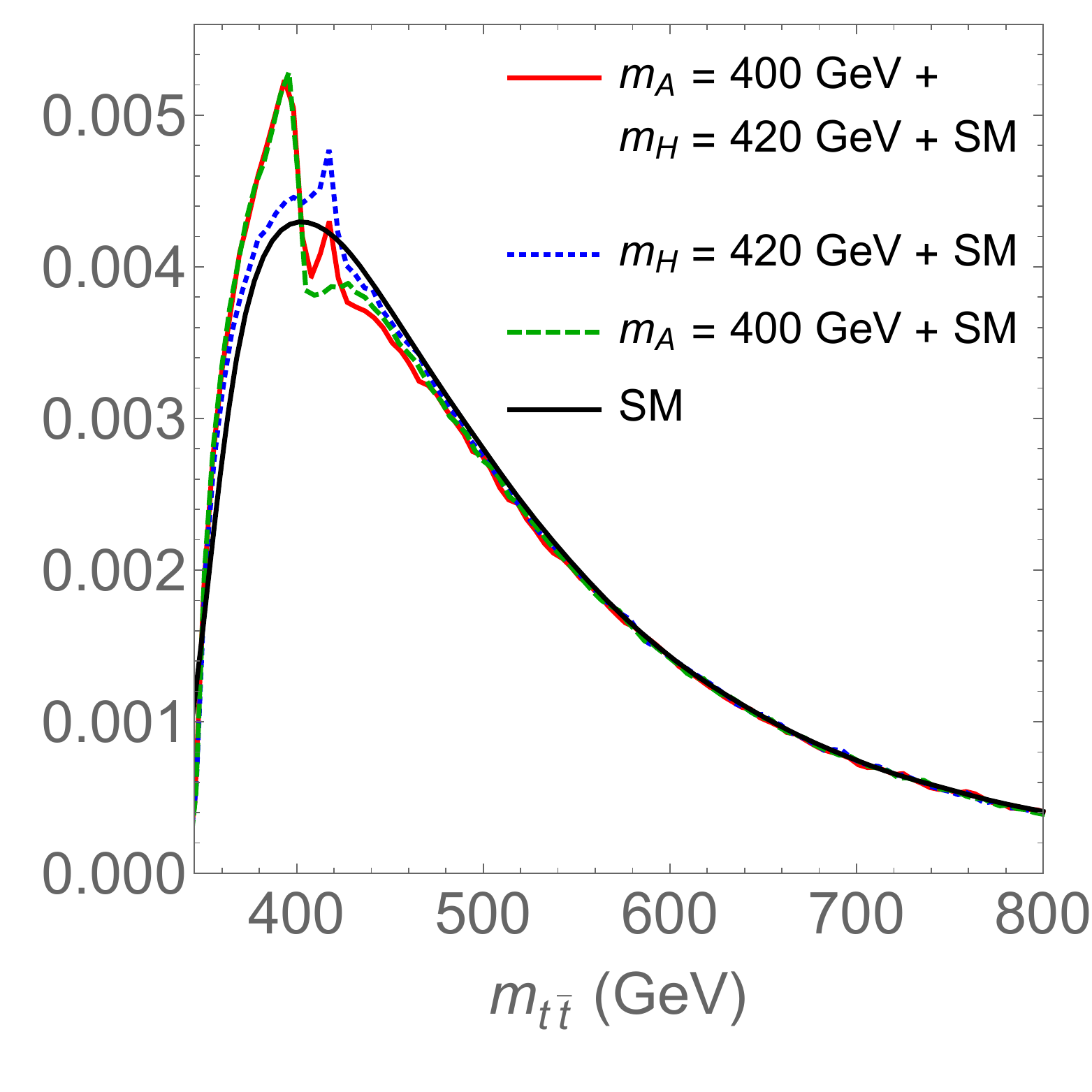}
\end{center}
\caption{The parton level invariant mass distribution for top pairs. The continuum SM background is shown in black solid, the interference between the pseudoscalar ($M_A=400$ GeV) only and the SM background in green dashed, the interference between the scalar only ($M_H=420$ GeV) and the SM background in blue dotted, and the SM background interfering with both the scalar and pseudoscalar in red solid. $\tan\beta$ is fixed to 1.}
\label{fig:2t_PL}
\end{figure}

We first validated our parton level simulation by verifying that the parton level events reproduce cross-sections in agreement with the theoretical expectation ~(see Figs.~2 \& 3 in Ref.~\cite{Dicus:1994bm}), with either one of $A$/$H$ or both together as expected in a MSSM scenario.  This can be seen from the invariant mass  distribution shown in Fig.~\ref{fig:2t_PL}, which we obtain using 1M parton level events each for the SM continuum (black solid line), scalar $H$~(dotted blue line, $m_H=420$ GeV) and pseudoscalar $A$~(dashed green line, $m_H=400$ GeV) separately interfered with the SM background.  We also show a typical MSSM scenario with $m_A$=400 GeV~(solid red line), where the heavy scalar $H$ will be split slightly in mass from $A$ at low values of $\tan\beta$ (taking a representative value of $m_H$=420 GeV).  We fix $\tan\beta=1$, so that $H$ and $A$ couple to the top quarks with a strength given by the top Yukawa. The expected dip and peak structure can be clearly seen in both the $H$ and $A$ interference separately. The slight off-set in the masses between $A$ and $H$ for the MSSM leads to a diminishing of the dip structure observed in the red solid line, obtained summing up the pseudoscalar and scalar contributions.  

\begin{figure}[tb]
\begin{center}
\begin{tabular}{ccc}
\includegraphics[width=0.45 \textwidth]{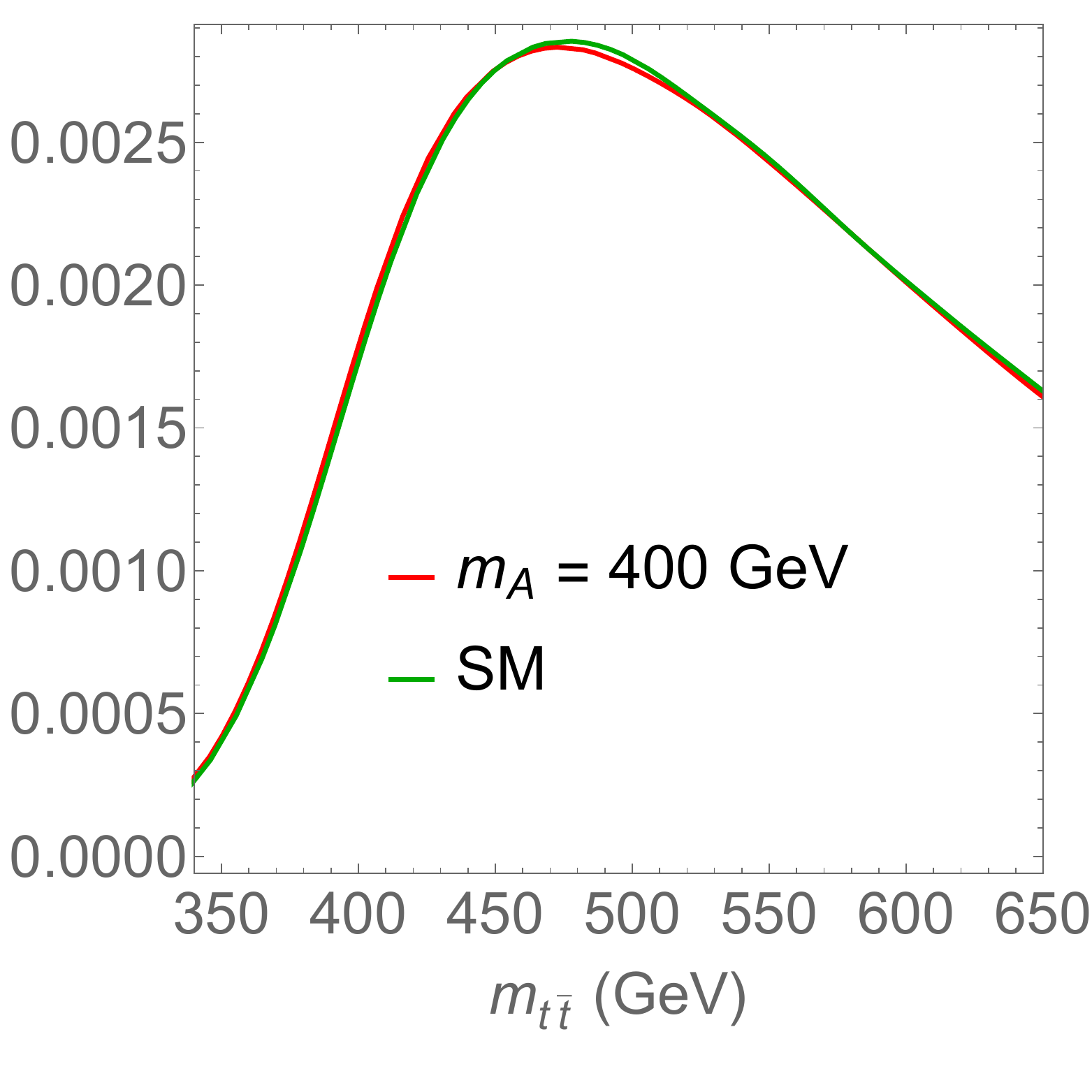} 
& \hspace{4mm} &
\includegraphics[width=0.49 \textwidth]{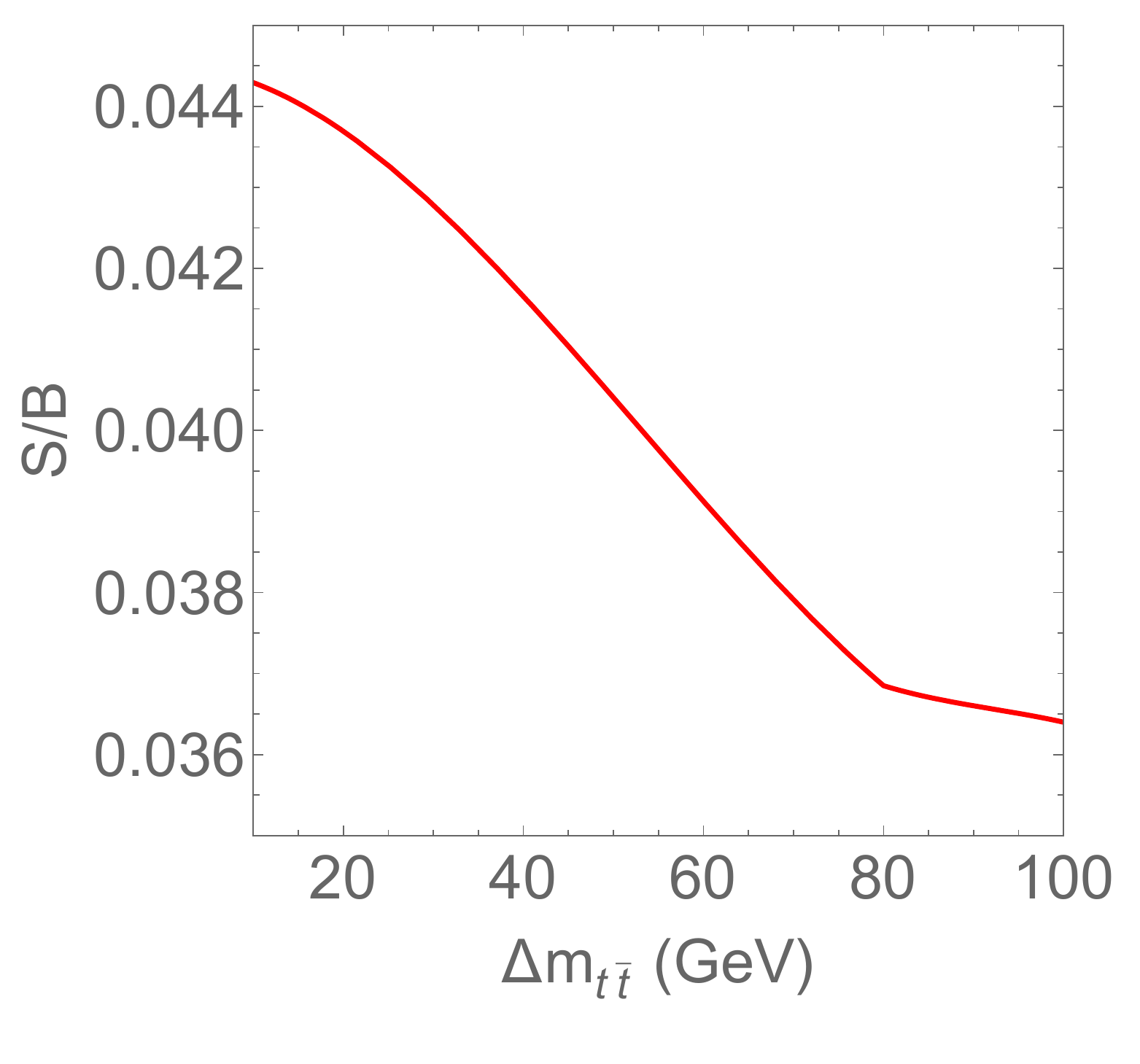} 
\end{tabular}
\end{center}
\caption{Left:  $m_{t \bar t}$ probability distribution for the SM background (green) and SM background interfered signal for $m_A = 400$ GeV, after showering, detector effects, and top reconstruction have been applied.  The distinctive peak-dip structure visible in Fig.~8 is washed out.  Right: $S/B$ as a function of the bin-width $\Delta m_{t\bar{t}}$. }
\label{fig:2t_PGS}
\end{figure}

The situation changes rather seriously once the tops are decayed, jets are showered, and smearing due to detector effects is taken into account. The tops are reconstructed using the method outlined in Appendix A.  The resulting distribution for the invariant mass of the two tops is shown in the left-hand panel of Fig.~\ref{fig:2t_PGS} for a pseudoscalar with a mass of 400 GeV interfering with the SM background (in red). The $t\bar t$ background is also shown in the figure (green solid line). For the plot, we generated 30M parton level events for signal and background. These substantial statistics are needed, since, after showering, detector simulation and top reconstruction less than a million events are left for the signal and background and differences between signal and background are only at the percent level. One can see that even without the presence of the $H$, which diminishes the peak-dip structure when split in mass from $A$, the interference with a 400 GeV Higgs (which has the largest signal)
is barely distinguishable from the SM only case. 
In the right-hand panel of Fig.~\ref{fig:2t_PGS}, we show $S/B$ as a function of $\Delta m_{t\bar{t}}$, where $S$ indicates the difference of number of events of signal plus background (including interference) and the number of events of the pure background. $\Delta m_{t\bar{t}}$ defines the width of the bin: $m_0\pm\Delta m_{t\bar{t}}/2$, where $m_0$ is fixed to $\sim 360$ GeV to maximize $S/B$.
As we can see from the figure, typically $S/B$ is on the order of $(3-4)\%$. We conclude, therefore, that 
the shape of the $m_{t\bar{t}}$ distribution must be very well-known before any conclusion could be drawn about the presence of additional Higgs bosons. In particular systematics would have to be very tightly controlled, which seems difficult at the level required. Even though the scale uncertainties in the SM only background and the signal interference approximately cancel when taking the ratio of the two, it was shown recently that NLO QCD effects can  easily be of the order of $\sim 5\%$~\cite{Bernreuther:2015fts}. It will be interesting to see if more precise theoretical studies will be able to bring this uncertainty on the $m_{t\bar t}$ distribution at the needed level in the future.

To conclude, the interference between $pp\to H/A\to t\bar t$ and the SM background strongly suggests that it will be quite challenging to set bounds on the $m_A-\tan\beta$ plane using a $t\bar t$ resonance search. The bounds will be much weaker than those extracted neglecting the interference effects and using LHC $t\bar t$ resonance searches performed either in the past or in the future to look for new gauge bosons decaying into two tops, $Z^\prime\to t\bar t$ (see, for example, Ref.~\cite{Djouadi:2015jea}).

%%%%%%%%%%%%%%%%%%%%%%%%%%%%%%%%%%%%%%%%%%%%%%%%%%%%%%%%%%%%
\section{Conclusions}\label{sec:conclusions}
%%%%%%%%%%%%%%%%%%%%%%%%%%%%%%%%%%%%%%%%%%%%%%%%%%%%%%%%%%%%%

\begin{figure}[tbph]
\begin{center}
\includegraphics[width=0.6 \textwidth]{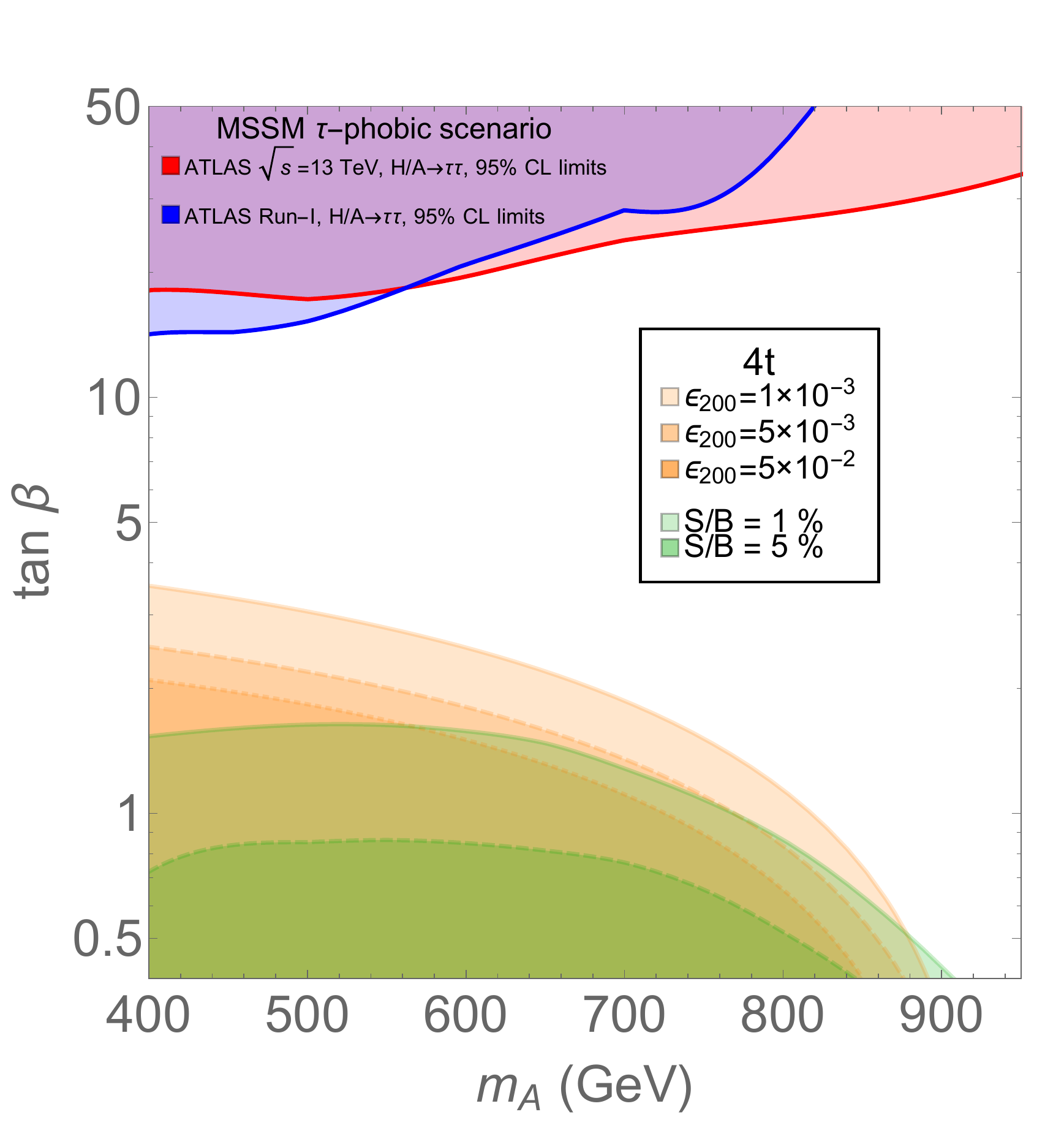}
\end{center}
\caption{Summary plot of the present bounds on heavy Higgs bosons at large values of $\tan\beta$ (red and blue regions from Ref.~\cite{ATLAStautau13}) and of the expected bounds at low values of $\tan\beta$ at the High Luminosity LHC14 (green and orange regions) from the $4t$ channel. The green and orange regions refer to our proposed Analysis (a.) and Analysis (b.), respectively. See Sec.~\ref{sec:4t} for more details. We interpret all bounds in terms of the $\tau$-phobic MSSM scenario.}
\label{fig:summary}
\end{figure}

We have studied prospects for the High Luminosity LHC to probe heavy Higgs bosons via heavy flavor final states, focusing on the $4 t$, $2b2t$ and $t \bar t$ signatures. These multi-top/multi-bottom final states are particularly interesting, since they are the ones with a sizable rate in Type II Two Higgs doublet Models in the (almost) alignment limit, as demanded by the SM-like properties of the discovered 125 GeV Higgs boson, for heavy Higgs bosons above the $t \bar t$ threshold, $m_A \gtrsim 350$ GeV and for moderate/small values of $\tan\beta$ to which the present LHC heavy Higgs searches are not sensitive (see the blue and red regions in Fig.~\ref{fig:summary} for the present LHC bounds on the $(m_A-\tan\beta)$ plane).

Focusing on the $\tau$-phobic MSSM scenario, we found that, at the High Luminosity LHC, the $4t$ signature arising from $pp\to t\bar t H/A,\, A/H\to t\bar t$ is promising for constraining heavy Higgs bosons with masses up to $\sim 750$ GeV, having a coupling to the top quarks which is the same as the SM top Yukawa.
In particular, for this signature, we compared the constraints from a single lepton plus multi-($b$) jet signature (Analysis (a.), see the shaded green regions in Fig.~\ref{fig:summary}) and from a multi-lepton (Analysis (b.) signature, see the shaded orange regions in Fig.~\ref{fig:summary}). For the single lepton signature, the main challenge is small signal-to-background ratio $S/B$: if the LHC will be able to reduce the systematic uncertainties to the level of $S/B\sim 1\%$, Higgs bosons as heavy as 750 GeV could be probed (again, if they couple to tops with a strength given by the SM top Yukawa). The reach completely degrades for $S/B=5\%$, for which heavy Higgs bosons could be probed only if they couple to tops with strength $\sim 1.3$ times larger than the SM top Yukawa. By contrast, we found that the multi-lepton analysis is not limited by small values of $S/B$, but will strongly depend on the specific fake rate efficiencies the LHC will be able to achieve. For fake rates similar to the ones at the 8 TeV LHC, heavy Higgses up to $\sim 750$ GeV with a SM top Yukawa coupling could be probed.

Unfortunately, both the $2b2t$ and $t \bar t$ signatures are dwarfed by the large $t \bar t$ background, and it will be hard for them to offer a meaningful reach in the $m_A-\tan\beta$ plane.  In particular, for the $2b2t$ signature, even when large enough $S/\sqrt{B}$ is obtained, $S/B$ is never as large as the few percent needed to overcome systematic uncertainties. This is due to the fact that, in spite of $\mathcal O(100)$ fb cross sections for $m_A=400$ GeV, this signature is very similar to the $t\bar t$ background: the two $b$ quarks produced in association with the Higgs boson have a low $p_T$ and a relatively sizable value of the pseudorapidity. We showed that, utilizing a higher $b$-tagging efficiency at large rapidity, as envisioned for the Phase II Upgrade, we will be able to set constraints on cross sections a factor of few smaller. Nevertheless, a more forward $b$-tagging will not be able to overcome the small $S/B$ ratio predicted by Type II Two Higgs doublet Models.

The $t \bar t$ signature also appears to offer, at first sight, good prospects for probing heavy Higgs bosons, due to relatively large cross sections of the order of a few pb for $m_A=400$ GeV. However, interference effects between the signal and the $t\bar t$ background seriously weaken the prospects of probing such heavy Higgs bosons, if the couplings are smaller than the top Yukawa coupling. To have prospects for setting bounds in the $m_A-\tan\beta$ plane, one will need to handle uncertainties on the $t\bar t$ background at the percent level or better.

The $4 t$ signature does, however, represent a genuine opportunity to constrain Two Higgs doublet Models in the ``wedge'' in the $m_A-\tan\beta$ plane at low $\tan \beta$ and above the $t \bar t$ threshold, where other searches are expected to be highly inefficient. 
We hope the LHC experiments will explore these signatures further.

\acknowledgments

We thank Marco Farina, Yuri Gershtein, Moira Gresham, Ben Hooberman, Fabio Maltoni, Michele Papucci, Brian Shuve and Carlos Wagner for useful conversations.  KZ is supported by contract DE-AC02-05CH11231.  N.R.S. thanks the hospitality of the
Aspen Center for Physics, which is supported by the National Science Foundation under
Grant No. PHYS-1066293.

\appendix

\section{Top Reconstruction}\label{Appendix}

The top reconstruction method we utilize follows Ref.~\cite{Gresham:2011dg}.  In order to obtain an accurate top reconstruction, the jet energies, mass,  $p_T$ and $\eta$ that feed into the top reconstruction method must be corrected for detector effects. This was accomplished by means of a pseudo-experiment, generating SM dijet events following Appendix 1a in Ref.~\cite{Gresham:2011fx}. For the 14 TeV LHC, we obtained the following
jet energy correction:
\begin{equation}
\frac{\Delta p_T}{p_{T_{\rm obs}}} = \frac{\Delta E}{E_{\rm obs}} =  \frac{16.81 + 6.83 \eta_{\rm obs}^2}{\sqrt{p_{T_{\rm obs}}^2 \cosh^2 \eta_{\rm obs} + m_{\rm obs}^2}} = \frac{\Delta m}{m_{\rm obs}}\; ; \qquad \Delta\eta=0\; , \label{jetcorr}
\end{equation}
where $m_{\rm obs}$ is a jet mass and all numbers are given in GeV.
The corresponding variance of the jet energy and angular parameters were found to be: 
\begin{equation}
\frac{\sigma_{p_T}}{p_{T_{\rm obs}}} = -0.036+\frac{2.7}{\sqrt{p_{T_{\rm obs}}}}, \qquad \sigma_{\eta} = \frac{1.06}{\sqrt{p_{T_{\rm obs}}}}\,. \label{sigmas}
\end{equation}

The procedure we utilize for top reconstruction is detailed in Appendices A and B of Ref.~\cite{Gresham:2011dg}. In particular, we find the missing neutrino momentum and fix combinatorics by minimizing the $\chi^2$ of our over-constrained system (for the definition of the $\chi^2$, see Appendices A and B of Ref.~\cite{Gresham:2011dg}). In the process of minimization over the missing neutrino momentum, we take into account the corrections to the jets in the {\tt PGS} events using  the functions given in Eq.~\ref{jetcorr}.  The variances given in Eq.~\ref{sigmas} are used as uncertainties in the measurement of the jet energies and angular resolution.   We verified that the $\chi^2$ distribution of the reconstructed top events is good, and reproduces well previous analyses utilizing the same method.

\begin{figure}[t]
\begin{center}
\begin{tabular}{ccc}
\includegraphics[width=0.45 \textwidth]{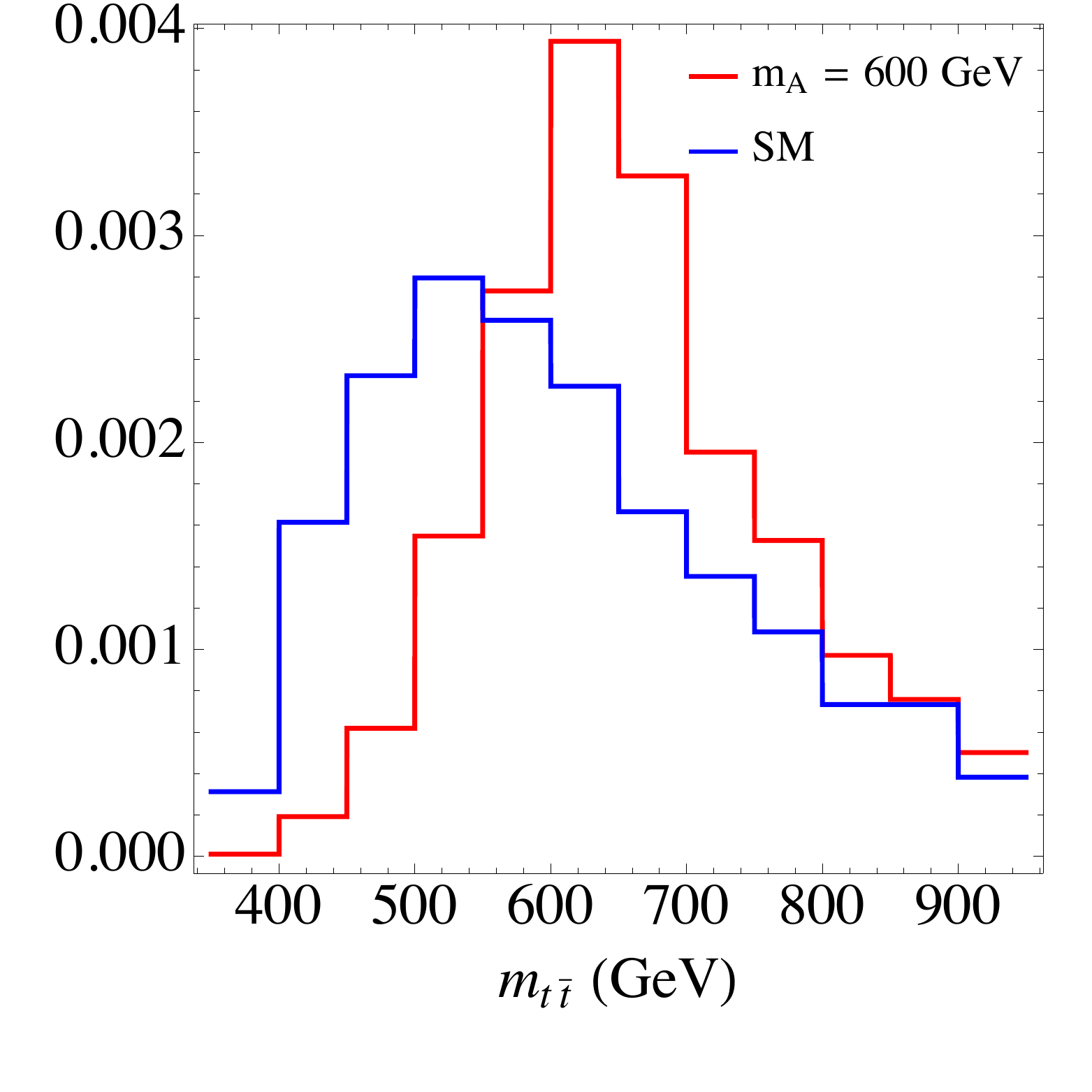}
& \hspace{4mm} &
\includegraphics[width=0.45\textwidth]{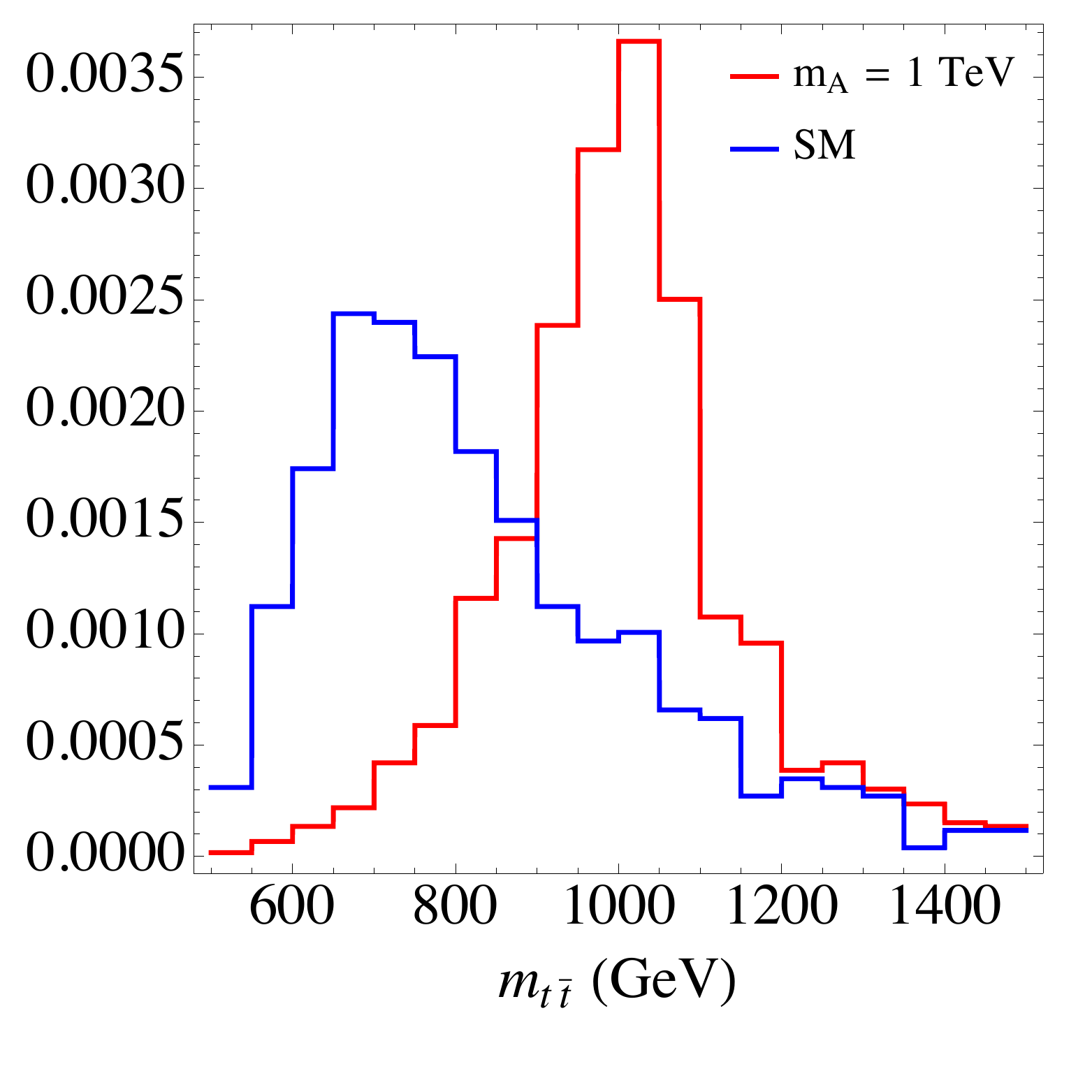}\\
\end{tabular}
 \includegraphics[width=0.45 \textwidth]{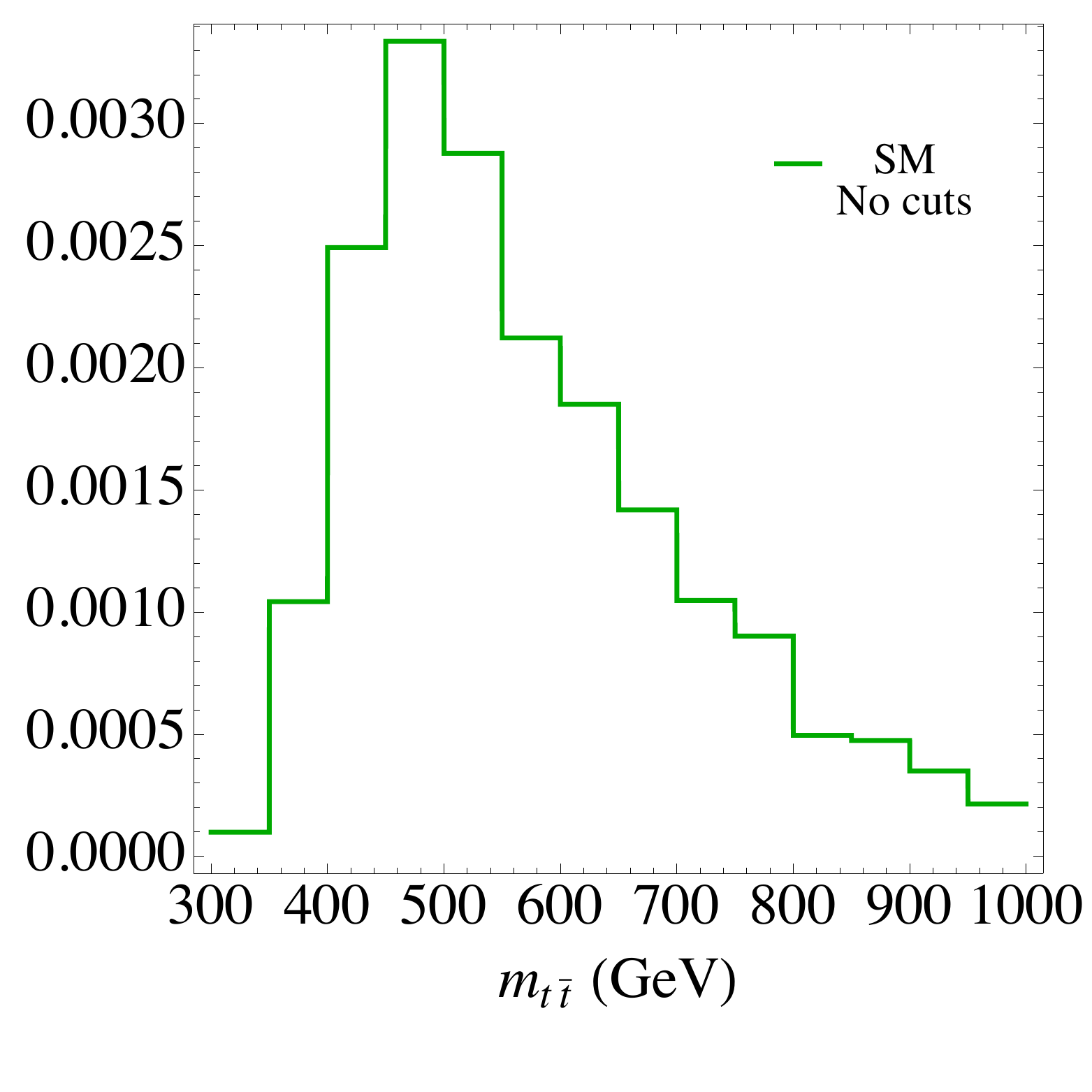}  
\end{center}
\caption{{\em Upper panels:} The probability distribution of the invariant mass of the reconstructed top pairs. Red: SM background, Blue: Signal events, Left panel: $m_A$=600 GeV, Right panel: $m_A$=1 TeV. Optimized cuts are imposed on both the SM background as well as the signal events before top reconstruction. {\em Lower panel:} The probability distribution for the invariant mass for the top pair reconstruction for SM events, without imposing any cuts.}
\label{fig:topreco600and1}
\end{figure}

We find that, despite the excellent performance of the top reconstruction algorithm employed, the invariant mass of the top pairs thus reconstructed did not give a very efficient discriminant between the signal and the background for the $2b2t$ case. For each value of $m_A$, we optimized the cuts as detailed in Sec.~\ref{sec:2t2b}. Events passing these cuts, both for the SM $t\bar t$ background and  signal, are fed into the top reconstruction algorithm. We find that the kinematic cuts employed in our analyses are very efficient, so much so that the top reconstruction employed on these events is unable to discriminate signal over background efficiently.  This can be seen from the sample distributions shown in Fig.~\ref{fig:topreco600and1} both for $m_A$ = 600 GeV (left) and $m_A$ = 1 TeV (right) and for the SM $t\bar t$+jets background (in blue). The distributions of the $t\bar t$+jets background in the left and right panel are obtained applying a set of  optimized cuts for $S/\sqrt{B}$ corresponding to these masses ($m_A=$ 600 GeV and $m_A=$ 1 TeV, respectively). The changes in efficiency due to top reconstruction are shown in Table~\ref{tab:topreco}. As can be seen from these numbers (and is visually apparent from Fig.~\ref{fig:topreco600and1}), top reconstruction does not  increase  $S/\sqrt{B}$. However, there may be some increase in $S/B$ (up to a factor of a few) for  masses of the order of $\mathcal{O}$(1) TeV, at the cost of reducing $S/\sqrt{B}$. 

\begin{table}
\begin{minipage}[b]{.95\textwidth}
  \centering
  \begin{tabular}{|c|c|c|c|  c  | c  |}
    \hline\hline
   & & \multicolumn{2} {|c|} {After Cuts} & \multicolumn{2}{|c|} {With Top Reconstruction}\\ 
    $m_A$  & $\sigma$ [fb] &$S/\sqrt{B}$& $S/B$& $\epsilon_{S/\sqrt{B}}$ & $\epsilon_{S/B}$\\
    \hline\hline
    600 GeV & 49 & 1.2 & $9\times 10^{-5}$ & 0.4 & 1.4 \\
    \hline
        1 TeV & 4& 0.2 & $3\times 10^{-5}$ & 0.5 & 3 \\
        \hline \hline
  \end{tabular}
\end{minipage}\qquad
\caption{Impact of top reconstruction on signal discrimination. The listed cross-section corresponds to $t_\beta=6$ in the $\tau$-phobic scenario for two heavy Higgs masses: $m_A=$ 600 GeV and 1 TeV. The given $S/\sqrt{B}$ and $S/B$ are computed after exemplary optimized cuts, assuming 3000 fb$^{-1}$ of data at LHC14. $\epsilon_{S/\sqrt{B}}$ and $\epsilon_{S/B}$ denote the impact of  top reconstruction (requiring $m_{t \bar t}$ be within 50 GeV of $m_A$) on $S/\sqrt{B}$ and $S/B$. The precise centering of the interval for the computation of the $\epsilon$ has not been optimized in the above, but it was checked that these numbers are representative of the change in efficiencies.}\label{tab:topreco}
\end{table}
For reference, the invariant mass distribution for the SM background, without imposing any cuts, is shown in the lower panel of Fig.~\ref{fig:topreco600and1}.

\bibliography{HeavyH}

%%%%%%%%%%%%%%%%%%%%%%%%%%%%%%%%%%%%%%%

\end{document}